\documentclass[preprint, 11pt]{elsarticle}

\usepackage[T1]{fontenc}
\usepackage[utf8]{inputenc}
\usepackage[hmargin=0.9in]{geometry}
\usepackage[babel,german=quotes]{csquotes}
\usepackage{subcaption}
\usepackage{graphicx}
\usepackage[colorlinks=true,citecolor=blue,urlcolor=blue]{hyperref}
\usepackage{caption}
\usepackage{amsmath}
\usepackage{amsthm}
\usepackage{amssymb}
\usepackage{bm, amsfonts}
\usepackage{xcolor,soul,framed}
\usepackage{fixmath}
\usepackage{tikz}
\usetikzlibrary{patterns, arrows.meta}
\usepackage{pgfmath}
\usetikzlibrary{3d,shapes,positioning}
\usepackage{pgffor}
\usepackage{subcaption}
\usepackage{caption}
\usepackage{comment}
\usepackage{bm}
\usepackage{bbm}
\usepackage{makecell}
\usepackage{titlesec}
\usepackage{multirow}
\usepackage{indentfirst}
\usepackage{mathtools}
\usepackage{float}
\usepackage[normalem]{ulem}
\useunder{\uline}{\ul}{}

\usepackage{booktabs} % For better table rules
\usepackage{siunitx}  % For aligning numbers in tables
\usepackage{adjustbox} % Alternative way to adjust the table size
\colorlet{shadecolor}{green}

\titleformat{\paragraph}
{\normalfont\normalsize\itshape}{\theparagraph}{1em}{}
\titlespacing*{\paragraph}
{0pt}{3.25ex plus 1ex minus .2ex}{1.5ex plus .2ex}

\newdefinition{remark}{Remark}

\newcommand{\beq}{\begin{equation}}
\newcommand{\eeq}{\end{equation}}

\titleformat{\subsection}{\bfseries\small}{\thesubsection}{1em}{}

\makeatletter
\def\ps@pprintTitle{%
   \let\@oddhead\@empty
   \let\@evenhead\@empty
   \def\@oddfoot{\centerline{\thepage}}%
   \def\@evenfoot{\centerline{\thepage}}%
}
\makeatother

\begin{document}

\begin{frontmatter}
  \title{Feasibility study for physics-informed direct numerical simulation describing particle suspension in high-loaded compartments of air-segmented flow}
\author{Otto Mierka$^1$, Raphael Münster$^{1,*}$, Henrik Julian Felix Bettin$^2$, Kerstin Wohlgemuth$^2$, Stefan Turek$^1$}
\ead{\{otto.mierka,raphael.muenster,stefan.turek,henrik.bettin,kerstin.wohlgemuth\}@tu-dortmund.de}
\cortext[cor1]{Corresponding author}

\address{$^1$ Institute of Applied Mathematics (LS III), TU Dortmund University\\ Vogelpothsweg 87, D-44227 Dortmund, Germany \\
$^2$ Laboratory of Plant and Process Design, TU Dortmund University\\
Emil-Figge-Straße 70, D-44227 Dortmund, Germany}

%\journal{Computers \& Chemical Engineering}
\journal{}

\begin{abstract}
The Archimedes Tube Crystallizer (ATC) employs air-segmented flow in coiled tubes to achieve narrow residence time distributions for continuous crystallization. Taylor and Dean vortices drive particle suspension in this system. However, one-way coupled models fail to capture the fluid–particle feedback that becomes critical at higher loadings. We present a particle-resolved Direct Numerical Simulation (DNS) framework based on a Finite Element–Fictitious Boundary Method with hard-contact modeling of particle interactions. Simulations of L-alanine suspensions across varying particle sizes, solid contents, and rotational speeds are validated against experimental side-view imaging. Three quantitative metrics—axial distribution, radial index, and vertical asymmetry—are introduced to classify suspension regimes. The DNS results reproduce the experimentally observed flow map zones (green, yellow, red/yellow, red) and resolve subtle transitions such as rear loading and loss of vertical symmetry. This feasibility study demonstrates that DNS can reliably predict dense suspension behavior and provides a mechanistic foundation for crystallizer design.
\end{abstract}


\end{frontmatter}

%%%
%%% Introduction
%%%

\section{Introduction} \label{intro}
Crystallization is a crucial process in the production of pharmaceuticals and chemicals, demanding precise control of critical quality attributes like particle size distribution (PSD) and crystal morphology to ensure efficacy \cite{Orehek.2020, Ma.2020, Eren.2023}. Achieving a narrow PSD is particularly important for downstream processing and product bioavailability \cite{Eren.2023, Chen.2011}. Continuous crystallization processes, offering benefits like consistency and improved product control compared to batch methods, are gaining attention \cite{Ma.2020, Eren.2023}. 
The Archimedes Tube Crystallizer (ATC) distinguishes itself through a coiled design and air-segmented flow. These features allow for flexible residence times and a narrow residence time distribution (RTD). By approximating ideal plug flow, the ATC creates the conditions necessary for uniform crystal growth and a narrow PSD \cite{Sonnenschein.2022, Sonnenschein.2021, Sonnenschein.2022b}.

Understanding the intricate hydrodynamics and suspension behavior of particles within the ATC is fundamental, as these phenomena significantly impact crucial aspects like agglomeration, breakage, attrition, and potential crystallizer blockage \cite{Sonnenschein.2022b, Nagy.2020}. Computational Fluid Dynamics (CFD) is an essential tool for probing these complex dynamics and simulating the behavior of liquid segments separated by an air phase.

Previous research on ATC hydrodynamics has applied various computational approaches. For instance, our preliminary study utilized a one-way coupled Lagrangian Particle Tracking (LPT) approach integrated with an FEM-based CFD solver \cite{cryst11121466}. We chose this method as a compromise between accuracy and cost for specific particle sizes and low volume fractions. While it accounts for particle motion under hydrodynamic drag, it neglects reciprocal momentum feedback. However, this approach is less accurate for dense suspensions where particles accumulate locally. This is because the Schiller–Naumann drag correlation is valid only up to certain volume fractions at moderate Reynolds numbers \cite{SchillerNaumann1935}. In general particle-laden flow simulations, coupled CFD–DEM approaches \cite{ligggths, KruggelEmden2008SelectionOA, KRUGGELEMDEN2007157} are popular because they can handle very large numbers of particles efficiently. They rely on unresolved particle–fluid coupling, in which fluid–particle interactions are approximated through correlations. This approximation can limit accuracy in dense suspensions compared to particle-resolved Direct Numerical Simulation (DNS).

For a deeper understanding of particle–fluid interactions within the crystallizer DNS offers significant advantages. DNS is a fully resolved approach that requires computational meshes significantly finer than the particle size, such that hydrodynamic forces and torques emerge directly from the solution of the Navier–Stokes equations. This contrasts with unresolved approaches (e.g. CFD–DEM), where forces are based on correlations and volume fractions, or one-way coupling, where particle feedback is ignored. DNS methods establish full two-way coupling between the fluid and particles. This is achieved through strategies such as the Immersed Boundary Method (IBM) \cite{peskin, uhlmann}, Fictitious Boundary Method (FBM) \cite{WanTurek2006a, fbm2012}, and Fictitious Domain Method (FDM) \cite{glowinski1, patankar2000}. These approaches allow for the accurate resolution of suspension-induced rheology and particle migration phenomena \cite{rod, review2012}.

The high fidelity of DNS makes it a promising tool for obtaining fundamental insights into ATC hydrodynamics and particle behavior. For example, DNS studies in helically coiled tubes have already indicated that particle residence time can be size-dependent due to secondary flow patterns \cite{yang2004}, and more generally, DNS has revealed suspension behaviors that bridge the gap between dilute and dense regimes \cite{PhysRevLett.107.188301, PhysRevLett.109.118305, PhysRevLett.129.078001}.
This level of detail is crucial for understanding complex phenomena that are sensitive to local conditions, such as the interplay between the hydrodynamic environment and particle behavior. Simpler models may fail to fully capture these nuances.

Despite its advantages, DNS comes with significant challenges, primarily related to computational cost. High-fidelity DNS simulations, especially when extended with Population Balance Equations (PBEs), demand substantial computational resources \cite{review2012}. Simulating particle counts above $10^4$ remains highly demanding, which is a major reason why unresolved approaches are still favored for process-scale studies. Nevertheless, the development of efficient multigrid FEM solvers \cite{WanTurek2007a, WanTurek2007b}, as well as advances in parallel DNS frameworks \cite{münster2025effectiveviscosityclosuresdense, münster2025chimeradomaindecompositionmethod}, are gradually expanding the feasible limits of DNS for multiparticle suspensions.

Given this background, the present work aims to explore the applicability and potential of the DNS approach for simulating particle behavior in the ATC. The goals of this work are twofold:
\begin{itemize}
    \item To validate the DNS approach against available experimental data and results from other validated computational methods for representative ATC operating conditions.
    \item To prove the feasibility of DNS for simulating a substantial number of particles relevant to ATC conditions, provide arguments that underline the benefits of DNS in ATC simulations and identifying strategies for handling larger particle counts in future, more comprehensive studies.
\end{itemize}

We present a fully resolved DNS framework that reveals detailed ATC dynamics, establishing a mechanistic foundation for high-fidelity crystallizer design.

\section{Mathematical Model Development} \label{mmd}
In this section we present the modelling of the fluid problems by the FEM-FBM using DNS. We focus on the key aspects of the method and the two-way coupling that was not present in earlier simulations where the LPT approach was employed.
\subsection{Mixed Fluid-Particle Flow Domain}
\label{sec:modeling}

We model the suspension of rigid particles in an incompressible Newtonian fluid using the  FEM–FBM approach \cite{WanTurek2006a, WanTurek2006b}. The coupled fluid–particle system is solved in a fixed background domain $\Omega_T$
\begin{equation}
\Omega_T = \Omega_f \cup \bigcup_{i=1}^N \Omega_i,
\end{equation}
where $\Omega_f$ is the fluid region and $\Omega_i$ denotes the subdomain occupied by the $i$-th rigid particle.

\subsection{Governing Equations}

The incompressible Navier–Stokes equations govern fluid motion in $\Omega_f$:
\begin{align}
\rho_f \left( \frac{\partial \mathbf{u}}{\partial t} + \mathbf{u} \cdot \nabla \mathbf{u} \right) - \nabla \cdot \boldsymbol{\sigma} &= \mathbf{0}, \\
\nabla \cdot \mathbf{u} &= 0,
\end{align}
where the Cauchy stress tensor for a Newtonian fluid is
\begin{equation}
\boldsymbol{\sigma} = -p I + \mu_f \left[\nabla \mathbf{u} + (\nabla \mathbf{u})^\top\right].
\end{equation}

Inside each particle domain $\Omega_i$, the velocity field satisfies a rigid-body motion constraint:
\begin{equation}
\label{eq::twowayA}
\mathbf{u}(\mathbf{x}) = \mathbf{U}_i + \boldsymbol{\omega}_i \times (\mathbf{x} - \mathbf{X}_i).
\end{equation}
According to the Newton–Euler equations each particle $i$ obeys
\begin{align}
M_i \frac{d \mathbf{U}_i}{dt} &= \Delta M_i \mathbf{g} + \mathbf{F}_i + \mathbf{F}_i^{\text{col}}, \\
\mathbf{I}_i \frac{d \boldsymbol{\omega}_i}{dt} + \boldsymbol{\omega}_i \times (\mathbf{I}_i \boldsymbol{\omega}_i) &= \mathbf{T}_i,
\end{align}
with effective buoyant mass $\Delta M_i = M_i - \rho_f |\Omega_i|$. Together with the boundary condition in eq. \ref{eq::twowayA} this allows us to set up the two-way coupling between the fluid and the solid particles.

\subsection{Fictitious Boundary Method}

In the FEM–FBM approach, the governing equations are extended to the full domain $\Omega_T$, enforcing rigid-body constraints explicitly in particle regions. This transforms the coupled fluid–solid problem into a single-domain formulation on a fixed mesh.

The hydrodynamic force $\mathbf{F}_i$ and torque $\mathbf{T}_i$ on each particle $i$ are computed via volume-integral approximations:
\begin{align}
\mathbf{F}_i &= -\int_{\Omega_T} \boldsymbol{\sigma} \cdot \nabla \alpha_i \, d\mathbf{x}, \\
\mathbf{T}_i &= -\int_{\Omega_T} (\mathbf{x} - \mathbf{X}_i) \times (\boldsymbol{\sigma} \cdot \nabla \alpha_i) \, d\mathbf{x},
\end{align}
where $\alpha_i$ is the indicator function of particle $i$ and $\mathbf{X}_i$ its center. This approach avoids surface integration and is particularly well-suited for structured meshes.

\subsection{Discretization And Numerical Solution Strategy}

We discretize the system in time using the strongly A-stable fractional-step-$\theta$ scheme \cite{Blasco_Codina_Huerta_1998,WanTurek2007a} and in space using hexahedral meshes with the $Q_2/P_1^{\mathrm{disc}}$ finite element pair for velocity and pressure, respectively. The rigid-body constraints are enforced strongly on all nodes within each particle region. A detailed derivation of the weak forms, time-stepping scheme, and force computation strategy is given in \cite{münster2025effectiveviscosityclosuresdense}. While DNS resolves hydrodynamic forces directly, near-contact interactions require additional modeling. We therefore extend the DNS formulation with a frictional hard-sphere contact model, which ensures realistic treatment of particle–particle and particle–wall collisions.

\section{Hard Contact Model for Rigid Particles}
\label{sec:lubrication}
We adopt a nonsmooth hard-contact formulation with Coulomb friction, which is the standard approach in rigid-body dynamics. Unlike soft-sphere DEM, this avoids introducing artificial spring–dashpot energies; dissipation arises solely from frictional laws. For details on the following variable definitions and operators, the reader is referred to Appendix~\ref{appendix:contact_symbols}. A more comprehensive description of the method is given in \cite{münster2025effectiveviscosityclosuresdense}.
Contacts are formulated as point-wise constraints and resolved as a linear complementarity problem based on the Delassus operator \cite{stewart_trinkle_1996, anitescu_potra_1997}:
\begin{equation}
\dot{\bm{g}}_i = \bm{W}_i \bm{p}_i + \bm{b}_i, \quad \bm{p}_i \in \mathcal{K}_i.
\end{equation}
Baumgarte stabilization \cite{baumgarte_1972} is applied to improve numerical stability in the normal direction.

The system of contacts is solved iteratively using the Projected Gauss–Seidel (PGS) method \cite{erleben_2004}:
\begin{equation}
\bm{p}_i^{k+1} = \Pi_{\mathcal{K}_i}\bigl(\bm{p}_i^k - \omega \bm{W}_{ii}^{-1}\dot{\bm{g}}_i \bigr).
\end{equation}
We implement both approximate decoupled and fully coupled friction-cone projection strategies \cite{hwangbo_lee_hutter_2018, anitescu_tasora_2010}.

Parallelization is achieved through domain decomposition, with subdomains exchanging ghost-body velocity data after each PGS sweep to maintain consistency across the simulation domain. This approach integrates naturally with the CFD solver's decomposition strategy and enables large-scale simulation of dense particle suspensions with frictional contact. As implementation base for the described hard contact model we use a fork of the Physics Engine (\textit{pe}) project \cite{EIBL201836, computation7010009} that has been extended with an efficient interface to our CFD solver.

Although the present study uses spherical particles, this choice reflects a modeling simplification rather than a limitation of the method. The underlying hard-contact formulation by Anitescu and collaborators \cite{anitescu_potra_1997} applies to arbitrary rigid-body geometries. We introduced the spherical approximation for two reasons. First, it ensures comparability with earlier LPT simulations \cite{cryst11121466}. Second, previous studies concluded that the spherical shape assumption does not interfere with flow regime identification. 

\subsection{Key Advantages of the combined DNS/Contact Model Method}
\label{sec:key_advantages_dns}

We contrast a Lagrangian point-particle (LPT) approach using the Schiller--Naumann (S--N) drag correlation with particle-resolved DNS for spherical particles in liquid--solid flows across dilute ($\phi \lesssim 0.01$) to moderately dense ($0.05 \le \phi < 0.2$) volume fractions $\phi$ and from viscous to moderately inertial particle Reynolds numbers \(Re_p\).

The S--N correlation for the drag coefficient of a single, isolated, rigid sphere in an unbounded fluid with common parameters shows a constant plateau \(C_D\approx 0.44\) for \(Re_p\gtrsim 10^3\) (pre-crisis inertial regime) \cite{CliftGraceWeber,BrownLawler2003,Achenbach1972}. The S--N model has no dependence on the local solids volume fraction \(\phi\), a property which our DNS approach clearly has, as shown in our prior work \cite{münster2025effectiveviscosityclosuresdense}. Even the low volume fractions used in this work show configurations where the local solid concentration is high (rear loading, stacking), a situation where models like S--N deviate from DNS solutions \cite{LaddHillKoch, Beetstra, TENNETI20111072} and typically under-predict drag. Furthermore, S--N provides drag only: lift and torque are not included. When shear/rotation matter, additional closures for lift and hydrodynamic torque are required (e.g.\ \cite{Saffman1965,HolzerSommerfeld2008,Zastawny2012}). In LPT/S–N, particles interact only via drag with a one-way or frozen carrier flow, so both collision-induced momentum transfer and lubrication effects from squeezed fluid films are neglected. DNS with sufficient grid resolution can capture these effects and apply them to the particles in close proximity as well as neighboring particles. When the geometric gap gets to small to be resolved by the grid then lubrication correction can be added for DNS \cite{münster2025effectiveviscosityclosuresdense}. LPT/S--N can also add lubrication models, but they again lack the back coupling to the fluid, the effects of a squeezed fluid film are not propagated to neighboring particles.
Particle agglomeration is a critical phenomenon in the ATC. To study it rigorously, one must resolve squeeze-film pressure, momentum redistribution, and the history of local clustering.

In particle-resolved DNS (immersed/fictitious-boundary/domain methods), the \emph{total hydrodynamic force and torque} on each particle are obtained by integrating the surface stresses, with no drag/lift/torque closures and with full two-way momentum coupling \cite{MittalIaccarino2005, Uhlmann2005, WanTurek2006b}. This inherently captures unsteady features (wake dynamics, vortex shedding, particle induces vortex modification) that influence instantaneous and mean forces. This provides an inherent ability to capture important flow features in the ATC.

The techniques described in the preceding sections explicitly resolve fluid--particle interactions by solving the incompressible Navier--Stokes equations with immersed boundary constraints for moving rigid particles. This approach accounts for full momentum exchange between the fluid and the dispersed phase, capturing feedback effects such as vortex modification, wake interactions, and near-wall accumulation. Inter-particle and particle--wall collisions are treated with a frictional hard-contact model, and hydrodynamic forces, including lubrication effects, are resolved without empirical closure laws. This procedure represents a significant advancement over our previous LPT--FEM approach \cite{cryst11121466}, enabling more accurate, realistic, and insightful investigations.
We validate the numerical framework using experimental data from the ATC. The following section details the material system and operating conditions used for this comparison.

\section{Archimedes Tube Crystallizer} \label{atc}
The ATC is a small-scale continuous crystallizer developed by \textsc{Sonnenschein and Wohlgemuth} \cite{Sonnenschein.2022}. 
Critical quality attributes such as product particle size and size distribution are vastly impacted by particle suspension within the ATC. If particles are insufficiently suspended, contact times between particles increase and agglomeration of particles is aggravated. Therefore, a narrow particles size distribution is directly linked to a low degree of agglomeration and a good particle suspension. \cite{Sonnenschein.2022b} 
\newline
To investigate the particle suspension, which is dependent on the operating parameters of the ATC, a model material system is chosen. 

\subsection{Model Material System}
\begin{figure} [H]
    \begin{subfigure}{0.5\textwidth}
        \includegraphics[width=1\linewidth]{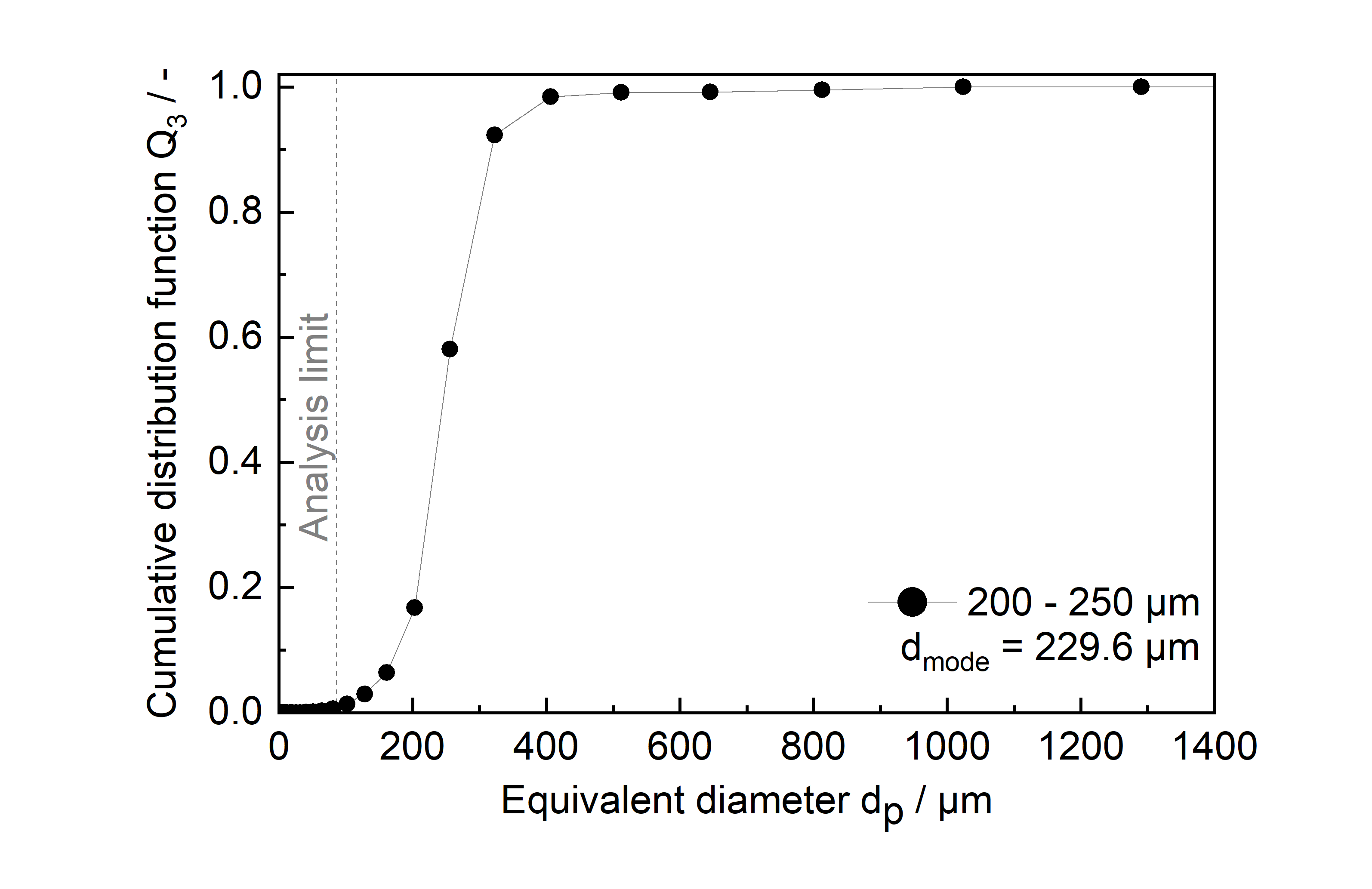}    
        \caption{Sieving fraction 200 - 250 µm}
        \label{fig:subim1}
    \end{subfigure}
     \begin{subfigure}{0.5\textwidth}
        \includegraphics[width=1\linewidth]{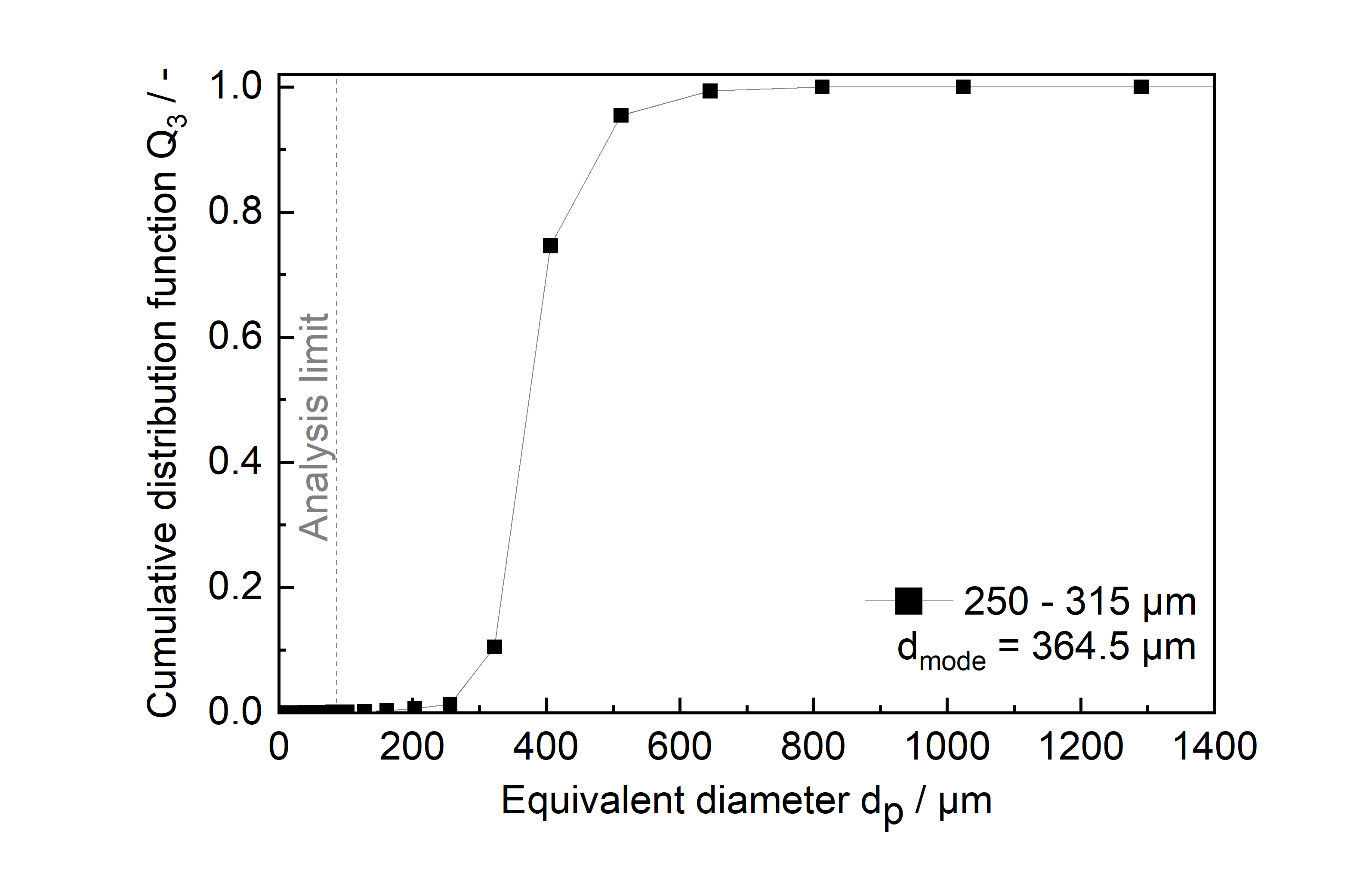}    
        \caption{Sieving fraction 250 - 315 µm}
        \label{fig:subim2}
    \end{subfigure}
    \caption{Cumulative particle size distributions $Q_3$ of the seed crystals obtained from the sieving fractions \newline 200 - 250 µm (a) and 250 - 315 (b). Additionally the respective mode diameter $d_{\text{mode}}$ is given.}
    \label{PSD}
\end{figure}

For experimental validation of the simulations, \textsc{l}-alanine (99.7 \% purity, Evonik Industries AG) in ultra pure water (18.2 \si{\mega\ohm\cm}) was selected as a model material system. For the liquid phase, a saturated \textsc{l}-alanine solution at ambient temperature was prepared according to equation \ref{csatala} and equilibrated for 48 hours \cite{Wohlgemuth2013modeling}. 
\begin{equation}\label{csatala}
    c^*(\vartheta)\Bigr[\si{\g_{ala}}\cdot\si{g_{sol}^{-1}}\Bigr] = 0.11238\cdot\exp{(9.0849\cdot 10^{-3} \cdot \vartheta[\si{\celsius}])}
\end{equation} 
The \textsc{l}-alanine seed crystals used as solid phase were prepared according to \textsc{Ostermann et al.} \cite{Ostermann2018Growthrates}. Subsequently, the obtained seed crystal fractions of 200 - 250 µm and 250 - 315 µm were analyzed via image analysis (QICPIC+LIXELL, Sympatec GmbH) to obtain a precise particle size distribution (Figure \ref{PSD}).

\subsection{Experimental Setup and Procedure}
\label{sec::experimenalSetup}
\begin{figure} [h]
    \centering
    \includegraphics[width=0.8\textwidth]{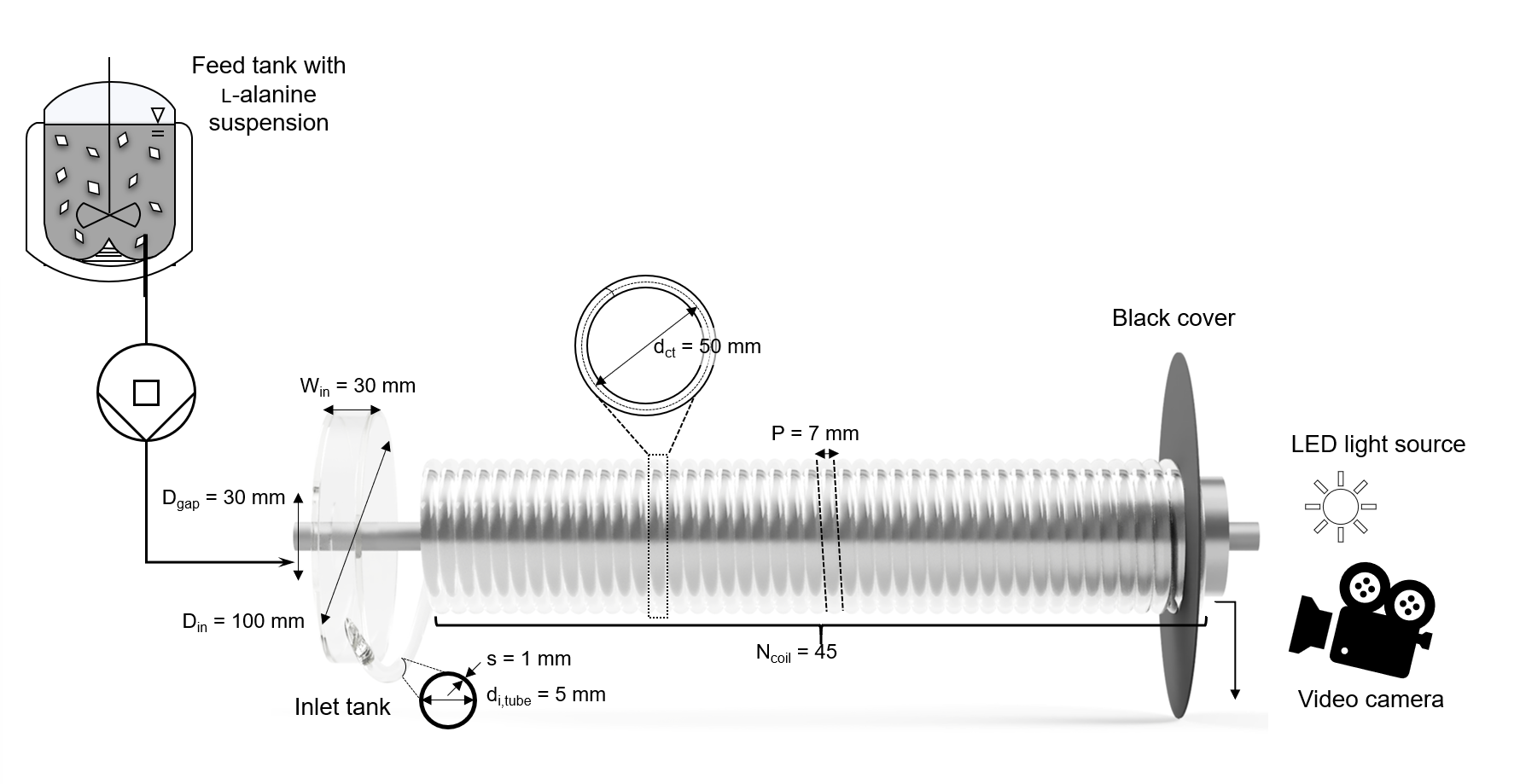}
    \caption{Schematic depiction of the setup used for the validation experiments. Additionally, the dimensions of the ATC are given. }
    \label{Setup}
\end{figure}

The experimental setup of the Archimedes Tube Crystallizer (ATC) and its periphery is shown in Figure \ref{Setup}. The feed tank, designed by \textsc{Lührmann et. al.}, provides the \textsc{l}-alanine suspension for the experiment \cite{Luhrmann2018continucleation}. Therefore, saturated \textsc{l}-alanine solution ($T_{\text{ambient}}$) is mixed with \textsc{l}-alanine seed crystals. For a homogeneous suspension, the liquid holdup of the feed tank is kept between 200-400 mL at a stirrer speed of 450 rpm. The suspension is transported via a peristaltic pump (Reglo-Digital MS-4, Ismatec) into the inlet tank of the ATC. The feed tank and the peristaltic pump are positioned above each other and higher than the ATC. Thereby, a blockage of the connecting tubes from sedimenting particles can be prevented. The structure of the silanized glass inlet tank leads to the formation of air-segmented liquid compartments due to a hole and the rotation of the apparatus. Each time the hole of the inlet tank is below the filling level, suspension flows into the connected silanized glass tube ($d_{\text{i,tube}} = 5$ mm), the actual ATC.
The operating parameters of the ATC are set by the  the rotational speed, filling degree, and the solid content inside of the suspension.
The rotational speed $n_{\text{ATC}}$ has a direct influence on the fluid flow inside the liquid compartments, where an increased $n_{\text{ATC}}$ results in an improved particle suspension \cite{Sonnenschein.2021}.
The filling degree $\varepsilon$ describes the fraction of one coil volume that is filled with suspension. The filling degree results out of $n_{\text{ATC}}$ and the feed volume flow of suspension. %For higher $\varepsilon$ around 0.5, the inlet tank does not empty with each rotation. Thereby, the residence time inside of the inlet tank broadens. Furthermore, the number of particles increases for constant solid contents, which leads to higher local concentrations of particles at subpar suspensions. --> There is no source for these claims.
The last modified parameter is the solid content $w_{\text{solid}}$ of the suspension.
Therefore, the segmented liquid compartments are transported through the coiled tubing toward the outlet with each rotation according to the selected $n_{\text{ATC}}$. At the end of the ATC, the particle suspension is evaluated by a video camera (Samsung NX300), which is positioned directly beneath the rotation axis of the ATC. For better contrast, the last coil is isolated via a 3D-printed black cover. An LED light source is used for illumination. After a steady state in the size and position of the liquid compartments is reached, the suspension is recorded.
\newline For evaluating the simulated suspension behavior using the model described above, the suspension is recorded at the following operating points stated in Table \ref{tab:exp_op_points}. Additionally, the fluid and particle properties on which the simulations are based are given.
\begin{table}[h]
    \caption{Operating points used in experiments (a). The filling degree is chosen as $\varepsilon$ = 0.25. Fluid and particle properties used in this work (b).}
    \label{tab:exp_op_points}
    \centering
    \begin{minipage}[t]{0.45\textwidth}
        \centering
        \subcaption{Operating points}
        \begin{tabular}{ccc}
            \toprule
            $n_\text{ATC}$ [rpm] & $d_{\text{mode}}$ [$\mu$m] & $w_\mathrm{solid}$ [wt\%]\\
            \midrule
            40 & 229.6 & 1.0 \\
            40 & 229.6 & 5.1 \\
            25 & 229.6 & 1.0 \\
            25 & 229.6 & 5.1 \\
            12 & 364.5 & 1.0 \\
            12 & 364.5 & 5.1 \\
            \bottomrule
        \end{tabular}
    \end{minipage}
        \begin{minipage}[t]{0.45\textwidth}
        \centering
        \subcaption{Physical Parameters}
        \label{tab:exp_op_points:properties}
        \begin{tabular}{ll}
        \toprule
        \multicolumn{2}{c}{Constant Physical Parameters} \\
        \midrule
        Fluid density & $\rho_f = 1043\,\mathrm{kg\,m^{-3}}$ \\
        Fluid viscosity & $\eta_f = 0.001\,\mathrm{Pa \cdot s}$ \\
        Particle density & $\rho_p = 1420\,\mathrm{kg\,m^{-3}}$ \\
        \bottomrule
      \end{tabular}
  \end{minipage}
\end{table}
To deepen our analysis, we will introduce three quantitative metrics in the next section that help us to enable objective classification of the suspension regimes beyond qualitative visual inspection.
\section{Quantitative Metrics for Flow Regime Classification in the ATC}
\label{sec:metricSection}
\begin{figure} [H]
    \centering
    \includegraphics[width=0.5\textwidth]{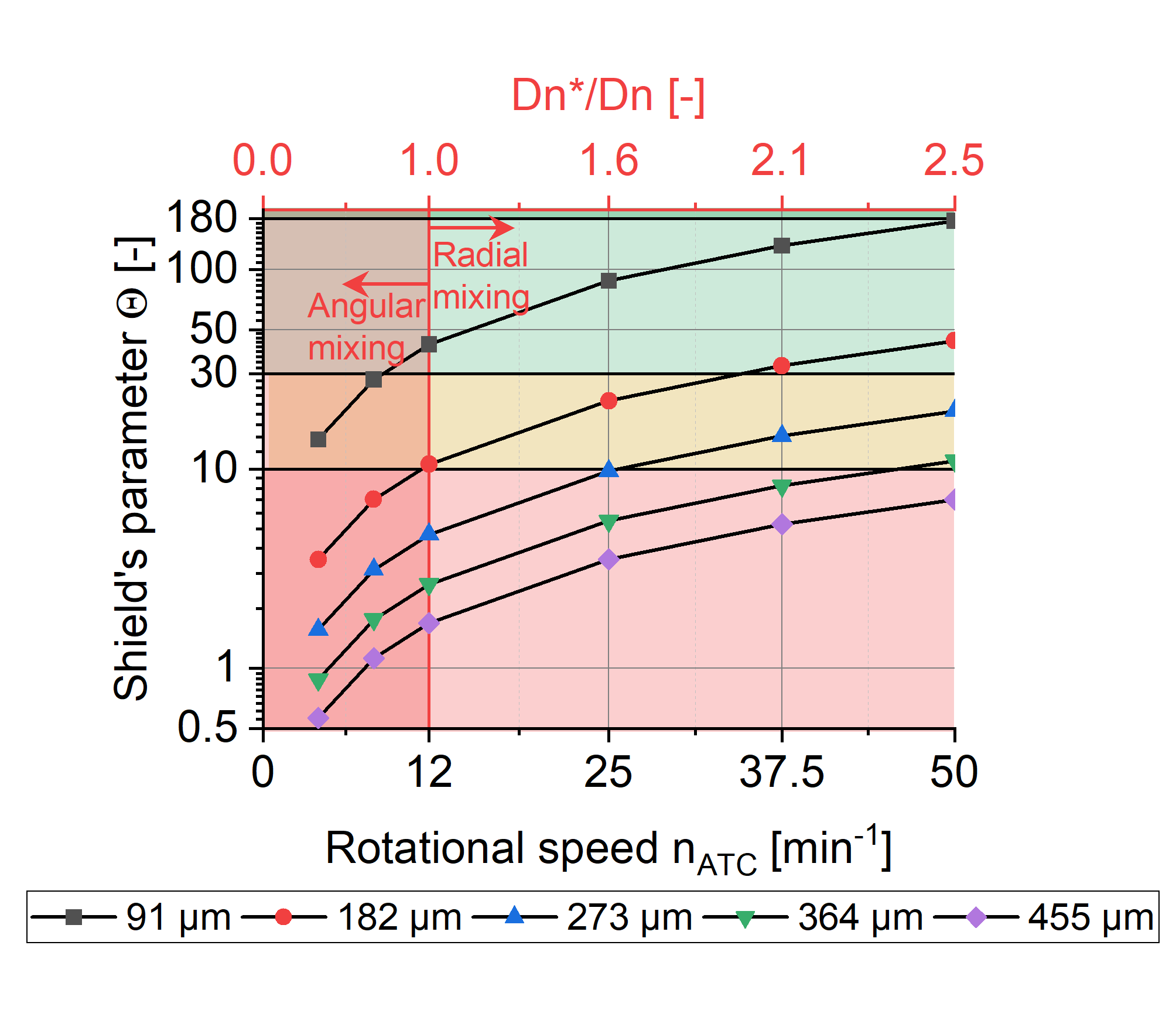}
    \caption{Regime classification flow map for the ATC. Reprinted from \cite{Sonnenschein.2021}. Copyright CC BY 4.0}
    \label{fig:flow_map_dct50}
\end{figure}

To complement visual comparisons between simulations and experiments in the Archimedes Tube Crystallizer (ATC), we introduce three quantitative metrics to characterize the internal suspension behavior of the solid phase. These are extracted from fully resolved direct numerical simulation (DNS) particle data and enable direct classification of the observed flow regime using the experimental flow map presented in~\cite{cryst11121466}, see also Figure \ref{fig:flow_map_dct50}.

%====
\subsection{Axial Particle Distribution \texorpdfstring{\(\overline{\phi}(s)\)}{phi(s)}}

The axial particle distribution is defined along the liquid compartment (commonly referred to as a `slug') centerline using the arc-length parameter \(s\), i.e.\ \(\overline{\phi}(s)\). This quantity describes the mean solid volume fraction along the axis of the slug and captures longitudinal segregation effects. 

For a liquid slug with curved centerline $\mathbf{c}(s)$ parameterized by arc length $s \in [0, L]$, the ideal axial particle distribution is defined as:
\begin{equation}
\overline{\phi}(s) = \frac{1}{|A(s)|} \int_{A(s)} \phi(\mathbf{r})\, dA ,
\end{equation}
where \(A(s)\) denotes the cross-sectional plane perpendicular to the centerline at arc length \(s\), and \(|A(s)|\) is its area. For the ATC tube geometry, this cross-sectional area is constant.

To evaluate \(\overline{\phi}(s)\) from discrete particle data in curved geometries, we employ an orthogonal projection method that reduces the 3D problem to 1D in the following way:

\begin{enumerate}
    \item Given the slug geometry, we extract the centerline $\mathbf{c}(s)$ where $s$ is the arc length parameter.
    
    \item For each particle at position $\mathbf{r}_i$, we find its projection onto the centerline by solving:
    \begin{equation}
    s_i = \arg\min_{s} \|\mathbf{r}_i - \mathbf{c}(s)\|^2 ,
    \end{equation}
    which yields the arc length coordinate $s_i$ corresponding to the point on the centerline closest to particle $i$.
    
    \item We normalize the arc length coordinates to obtain:
    \begin{equation}
    \tilde{s}_i = \frac{s_i}{L} \in [0,1] ,
    \end{equation}
    where $L$ is the total centerline length.
\end{enumerate}

With the normalized coordinates $\{\tilde{s}_i\}_{i=1}^{N_p}$, we approximate \(\overline{\phi}(s)\) using a histogram approach:

\begin{enumerate}
    \item \textbf{Binning}: Divide the normalized domain into $N_{\text{bins}}$ equal-width bins:
    \[
    B_k = \left[\frac{k-1}{N_{\text{bins}}}, \frac{k}{N_{\text{bins}}}\right), \quad k = 1, 2, \ldots, N_{\text{bins}} .
    \]
    
    \item \textbf{Particle Counting}: Count particles in each bin:
    \[
    n_k = \sum_{i=1}^{N_p} \chi_{B_k}(\tilde{s}_i) ,
    \]
    where $\chi_{B_k}$ is the indicator function of bin $B_k$.
    
    \item \textbf{Concentration Ratio}: Compute the dimensionless concentration ratio:
    \[
    \overline{\phi}_k = \frac{n_k}{n_{\text{uniform}}} = \frac{n_k N_{\text{bins}}}{N_p} .
    \]
\end{enumerate}

The histogram evaluation approximates the continuous integral through the following correspondences:

\begin{itemize}
    \item In the continuous formulation, \(\overline{\phi}(s)\) integrates particle volume fraction over each cross-section. In the discrete case, we count particles whose centers lie within a finite arc-length interval $\Delta s = L/N_{\text{bins}}$.
    
    \item The orthogonal projection implicitly performs the cross-sectional integration by mapping all particles in a cross-sectional vicinity to the same arc length coordinate:
    \[
    \frac{1}{|A(s)|} \int_{A(s)} \phi(\mathbf{r})\, dA \;\approx\; \frac{n_k}{V_k/v_p} ,
    \]
    where $V_k = |A(s)| \, \Delta s$ is the volume of bin $k$ and $v_p$ is the particle volume.
    
    \item By using the concentration ratio \(\overline{\phi}_k\), we obtain a dimensionless quantity that represents
    \[
    \overline{\phi}(s) \approx \overline{\phi}_k \cdot \phi_{\text{bulk}}, \quad s \in B_k ,
    \]
    where \(\phi_{\text{bulk}} = N_p v_p / V_{\text{total}}\) is the bulk volume fraction. This normalization ensures that averaging over all bins recovers the bulk value.
\end{itemize}

%====
Peaks in \(\overline{\phi}(s)\) near the rear of the slug are characteristic of gravitational settling and low Dean number regimes (``red zone''), while flatter profiles indicate increasing longitudinal homogenization associated with vortex-induced mixing (``green zone'').

\subsection{Radial Distribution Index \texorpdfstring{\(I_r\)}{Ir}}

The radial distribution index \(I_r \in [0,1]\) quantifies how uniformly particles are distributed within the cross-section of the slug.

Given a particle position \(\vec{p}_i\), we:
\begin{itemize}
  \item Project it onto the nearest point \(\vec{c}(s^*)\) on the ATC centerline.
  \item Compute the cross-sectional distance \(r_i = \|\vec{p}_i - \vec{c}(s^*) - ((\vec{p}_i - \vec{c}(s^*)) \cdot \vec{T}) \vec{T}\|\).
\end{itemize}

All radii \(\{r_i\}\) are binned into \(N\) equal-area radial shells. Let \(\phi_j\) be the particle count in shell \(j\), and \(\phi_\text{bulk}\) the mean count per shell. Then:
\begin{equation}
  I_r = 1 - \frac{1}{2 \phi_\text{bulk}} \sum_{j=1}^{N} \left| \phi_j - \phi_\text{bulk} \right|.
\end{equation}
Low values (\(I_r \ll 1\)) indicate wall accumulation; \(I_r \to 1\) indicates a uniform radial spread typical of well-mixed regimes.

\subsection{Vertical Asymmetry Index \texorpdfstring{\(A_y\)}{Ay}}

The vertical asymmetry index \(A_y \in [-1,1]\) measures the net imbalance of particle positions across the vertical mid-plane of the slug. It is defined using the same local Frenet frame as above.

Given the binormal vector \(\vec{B}\), which serves as the local ``upward'' direction, we define:
\begin{equation}
  \text{sign}_i = \operatorname{sign} \left( (\vec{p}_i - \vec{c}(s^*)) \cdot \vec{B} \right)
\end{equation}
to classify particles as above or below the vertical mid-plane. The index is then computed as:
\begin{equation}
  A_y = \frac{n_{\text{upper}} - n_{\text{lower}}}{n_{\text{upper}} + n_{\text{lower}}},
\end{equation}
where \(n_{\text{upper}}\) and \(n_{\text{lower}}\) are the total particle counts in each half. Values \(A_y < 0\) indicate bottom-heavy configurations dominated by gravity, while \(A_y \approx 0\) reflects vertical symmetry due to balanced Dean vortex structures.

\subsection{Usage in Regime Classification}

Each of the three metrics provides complementary information:

\begin{itemize}
  \item \textbf{Axial profile \(\overline{\phi}(s)\)} captures rear-loading vs uniformity.
  \item \textbf{Radial index \(I_r\)} quantifies in-plane mixing or wall accumulation.
  \item \textbf{Vertical asymmetry \(A_y\)} distinguishes gravitational vs vortex-driven distribution.
\end{itemize}

These metrics can be time-averaged and compared across operating conditions to classify simulations according to the ATC flow map:

\begin{table}[ht]
\centering
\caption{Linking the color coded flow regimes (see Figure \ref{fig:flow_map_dct50} from publication ~\cite{cryst11121466} for details) to metric value ranges.}
\label{tab:flowmap_metrics}
\begin{tabular}{lccc}
\toprule
\textbf{Flow Regime (map zone)} & \(\overline{\phi}(s)\) & \(I_r\) & \(A_y\) \\
\midrule
Red (gravitational)   & Peak at rear          & $\lesssim 0.5$   & $ \to -1$ \\
Yellow                & Moderate spread       & $0.5 < I_r < 0.8$ & $\ll 0$ \\
Green (fully mixed)   & Nearly flat profile          & $\to 1$           & $ \to 0$ \\
\bottomrule
\end{tabular}
\end{table}

Together, \(\overline{\phi}(s)\), \(I_r\), and \(A_y\) form a robust feature set for automated regime identification and comparison between DNS predictions and experimental observations in the ATC system. Furthermore, these metrics can be seen as a fine-grained variant of the goodness of suspension (GoS) metric defined in a prior publication \cite{TERMUHLEN2021116771}, used for quantitave evaluation of particle suspension in a slug
flow crystallizer.
Having established an upgraded DNS framework capable of handling representative ATC operating points, we now evaluate the results qualitatively—through direct side-view comparison with experiments—and quantitatively, using the metrics defined above.

\section{Results} \label{results}

In this study, a single liquid slug segment within the ATC is simulated to investigate particle suspension behavior under representative operating conditions. The ATC's rotation is modeled by applying boundary conditions directly on the surface of the liquid slug mesh. Specifically, Dirichlet boundary conditions consistent with the rotational speed of the apparatus are imposed on the tube walls, while free-slip conditions are applied at the spherical gas–liquid interfaces at the slug ends. This setup replicates the internal recirculation patterns typical of Taylor vortices while incorporating the centrifugal effects that generate Dean vortices.
Particle transport within the slug is resolved using the two-way coupled DNS approach (see \ref{sec:modeling}). 

The ATC geometry simulated corresponds to a coiled tube with a diameter of $d_{ct} = 50\,\text{mm}$, consistent with the experimental setup described in prior validation work (see also Figure \ref{Setup}). 
Simulations are performed across a range of ATC rotational speeds ($n_\mathrm{ATC}$) to match experimental conditions designed to generate different flow regimes. The configuration of the cases is summarized in Table \ref{tab:allcases}, while a list of fluid and particle properties can be found in Table \ref{tab:exp_op_points:properties}.

\begin{table}[H]
    \centering
    \caption{Summary of DNS simulation setup and cases, ordered by flow map zone. All simulations use \textsc{l}-alanine particles suspended in an aqueous solution. Cases are defined by varying rotational speed, particle diameter, and solid content at a constant filling degree of $\varepsilon$ = 0.25.}
    \label{tab:allcases}
    \begin{tabular}{ccccc}
        \toprule
        Case ID & $n_\text{ATC}$ [rpm] & $d_p$ [$\mu$m] & $w_\mathrm{solid}$ [wt\%] & Flow Map Zone \\
        \midrule
        C1 & 40 & 182 & 1.0 & green \\
        C2 & 40 & 182 & 5.1 & green \\
        C3 & 40 & 273 & 1.0 & yellow \\
        C4 & 40 & 273 & 5.1 & yellow \\
        C5 & 25 & 273 & 1.0 & red/yellow \\
        C6 & 25 & 273 & 5.1 & red/yellow \\
        C7 & 12 & 364 & 1.0 & red \\
        C8 & 12 & 364 & 5.1 & red \\
        \bottomrule
    \end{tabular}
\end{table}

\begin{figure}[H]

  \centering
  \includegraphics[width=\textwidth]{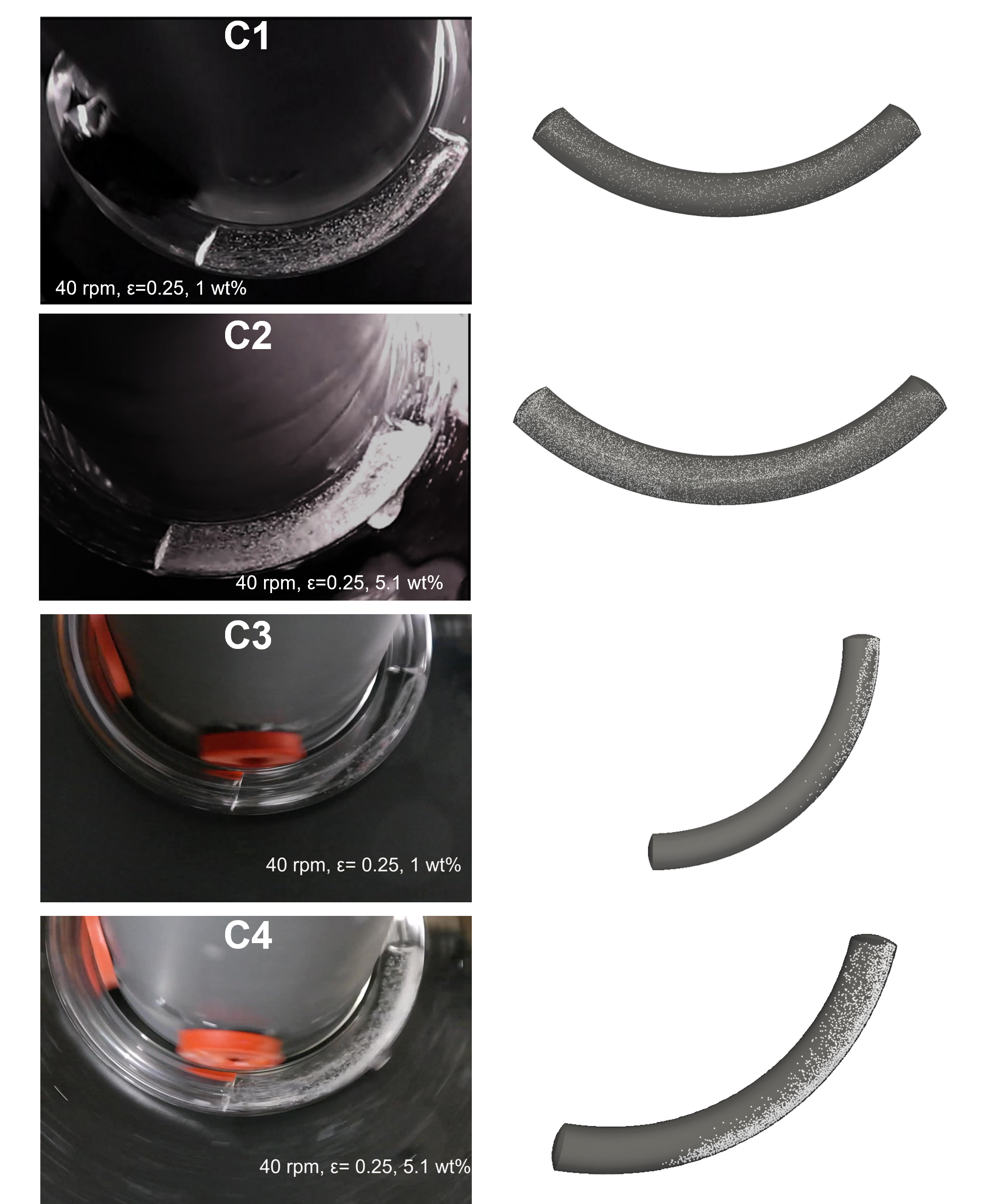}

  \caption{Comparison for cases C1, C2, C3, and C4. C1 and C2 reprinted from \cite{Sonnenschein.2021}. Copyright CC BY 4.0}
  \label{fig:image_comp_all}
\end{figure}

First, the simulation results are compared directly to experimental observations using side-view imaging (see Figure \ref{fig:image_comp_all}) of particle distribution within the ATC, as explained in section \ref{sec::experimenalSetup}. Secondly, the quantitative evaluation using the metrics defined in section \ref{sec:metricSection} validate the simulation approach, demonstrating that the two-way coupled DNS framework reliably predicts suspension states across the experimentally observed flow regimes. This integrated validation ensures that the simulated flow fields and suspension patterns correspond realistically to those observed in the laboratory, supporting the predictive use of the model for design and optimization of ATC operation. 

\subsection{Qualitative Comparison with Experimental Data}

\begin{figure}[H]
  \centering
  \includegraphics[width=\textwidth]{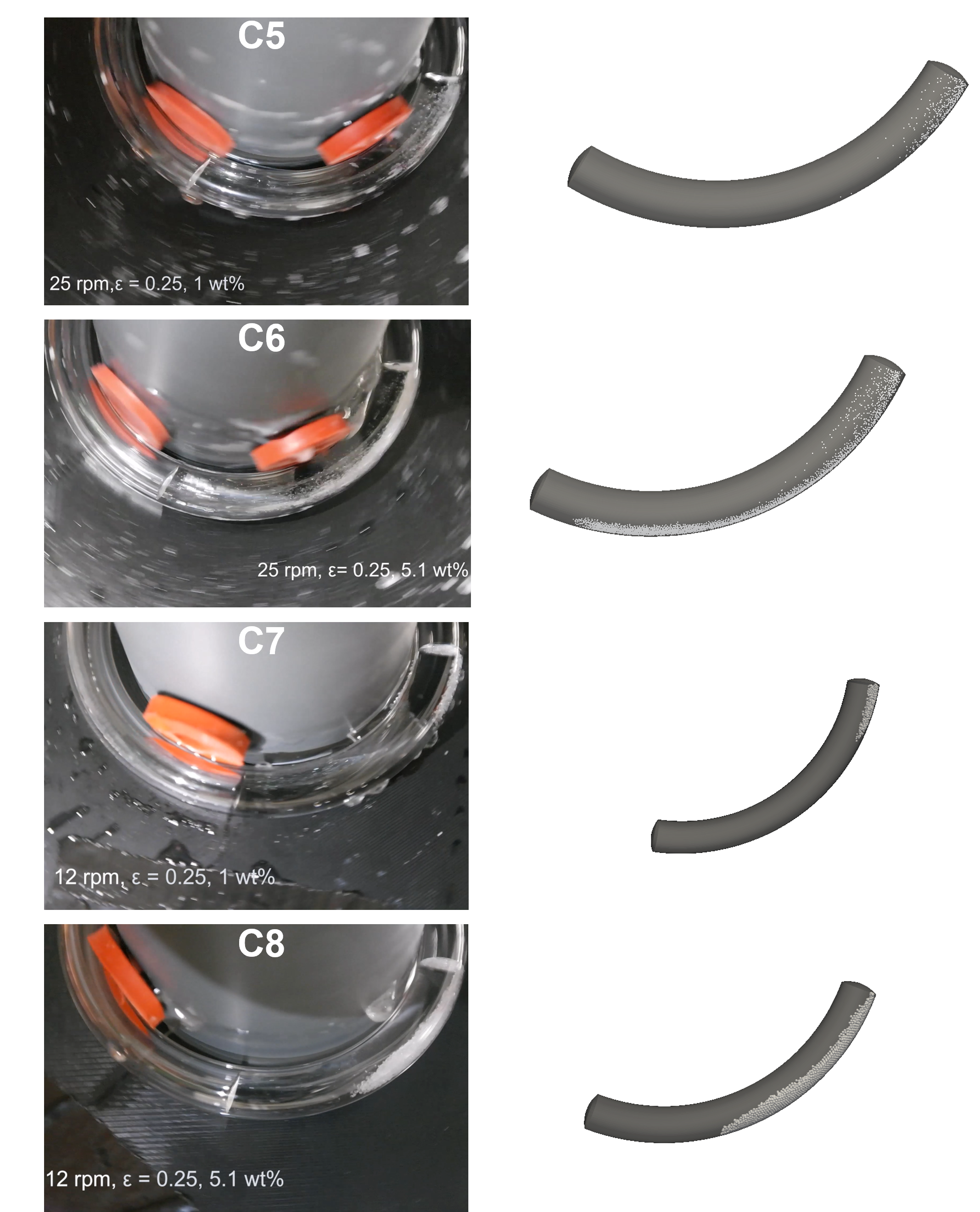}

  \caption{Comparison for cases C5, C6, C7, and C8.}
\end{figure}

\subsection{Metric-based Quantitative Analysis}

% First four cases
\begin{figure}[H]
    \centering
    \begin{subfigure}[b]{0.48\textwidth}
        \includegraphics[width=\textwidth]{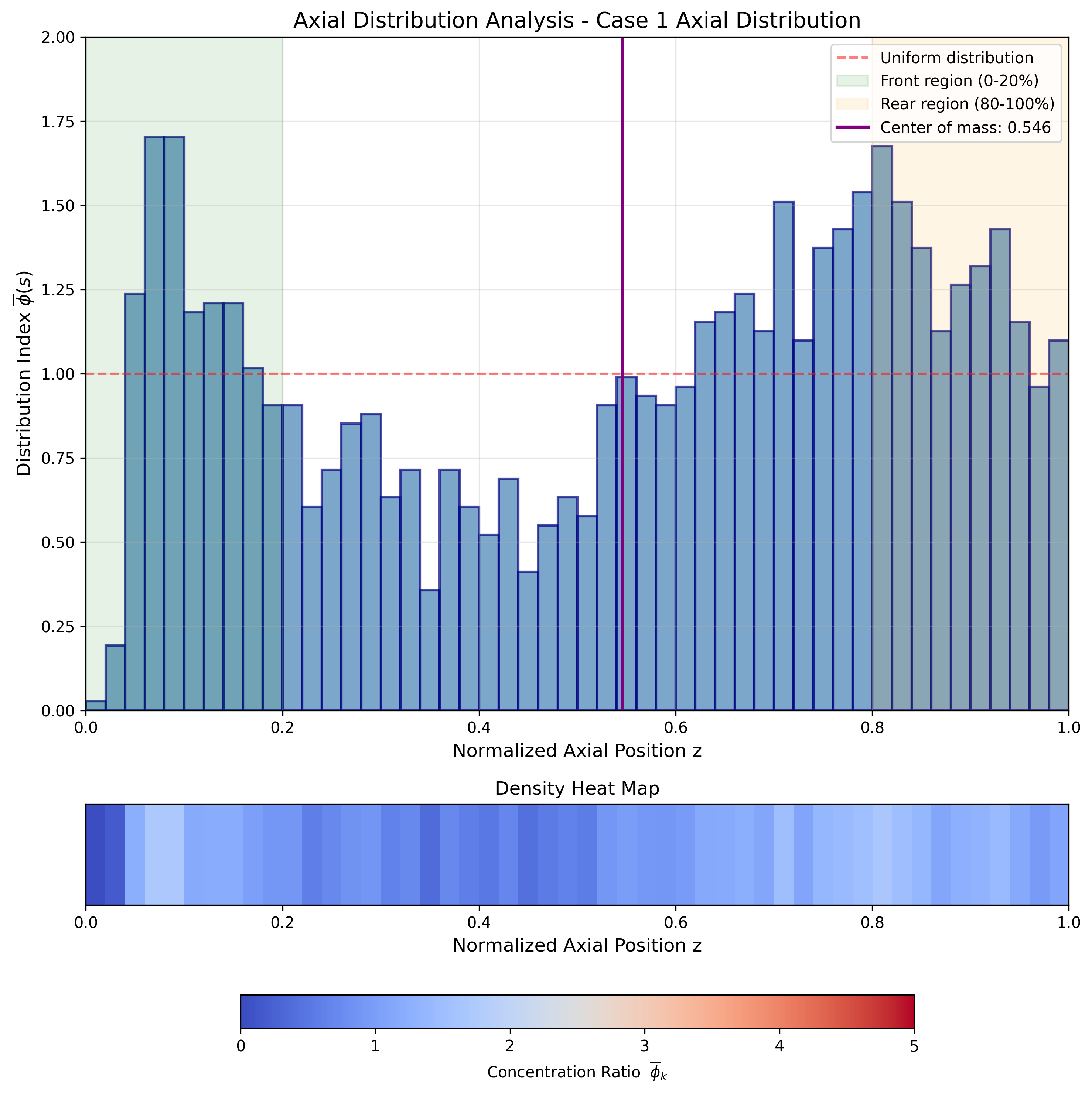}
        \caption{Case C1}
    \end{subfigure}\hfill
    \begin{subfigure}[b]{0.48\textwidth}
        \includegraphics[width=\textwidth]{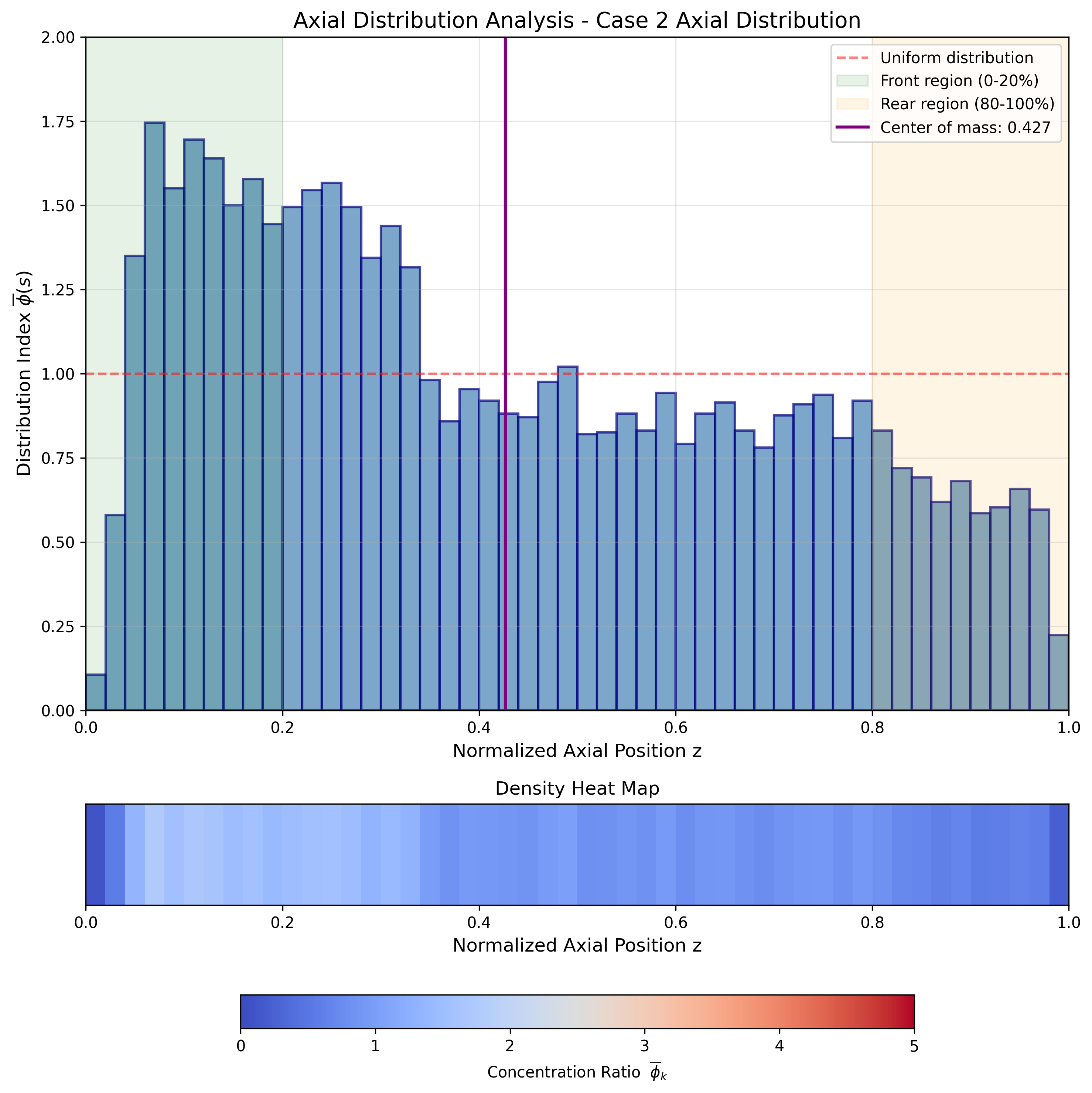}
        \caption{Case C2}
    \end{subfigure}

    \vspace{1em}

    \begin{subfigure}[b]{0.48\textwidth}
        \includegraphics[width=\textwidth]{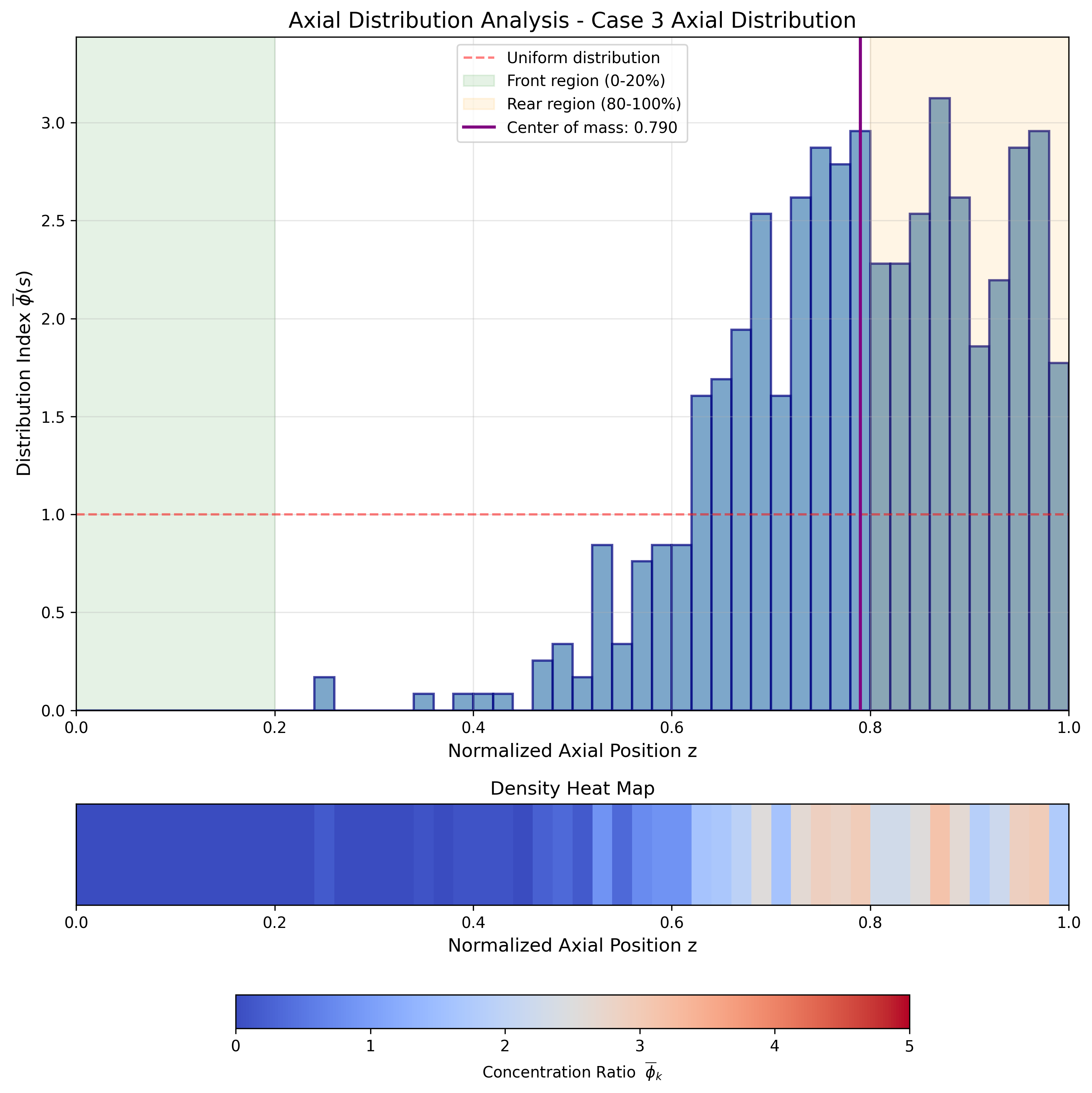}
        \caption{Case C3}
    \end{subfigure}\hfill
    \begin{subfigure}[b]{0.48\textwidth}
        \includegraphics[width=\textwidth]{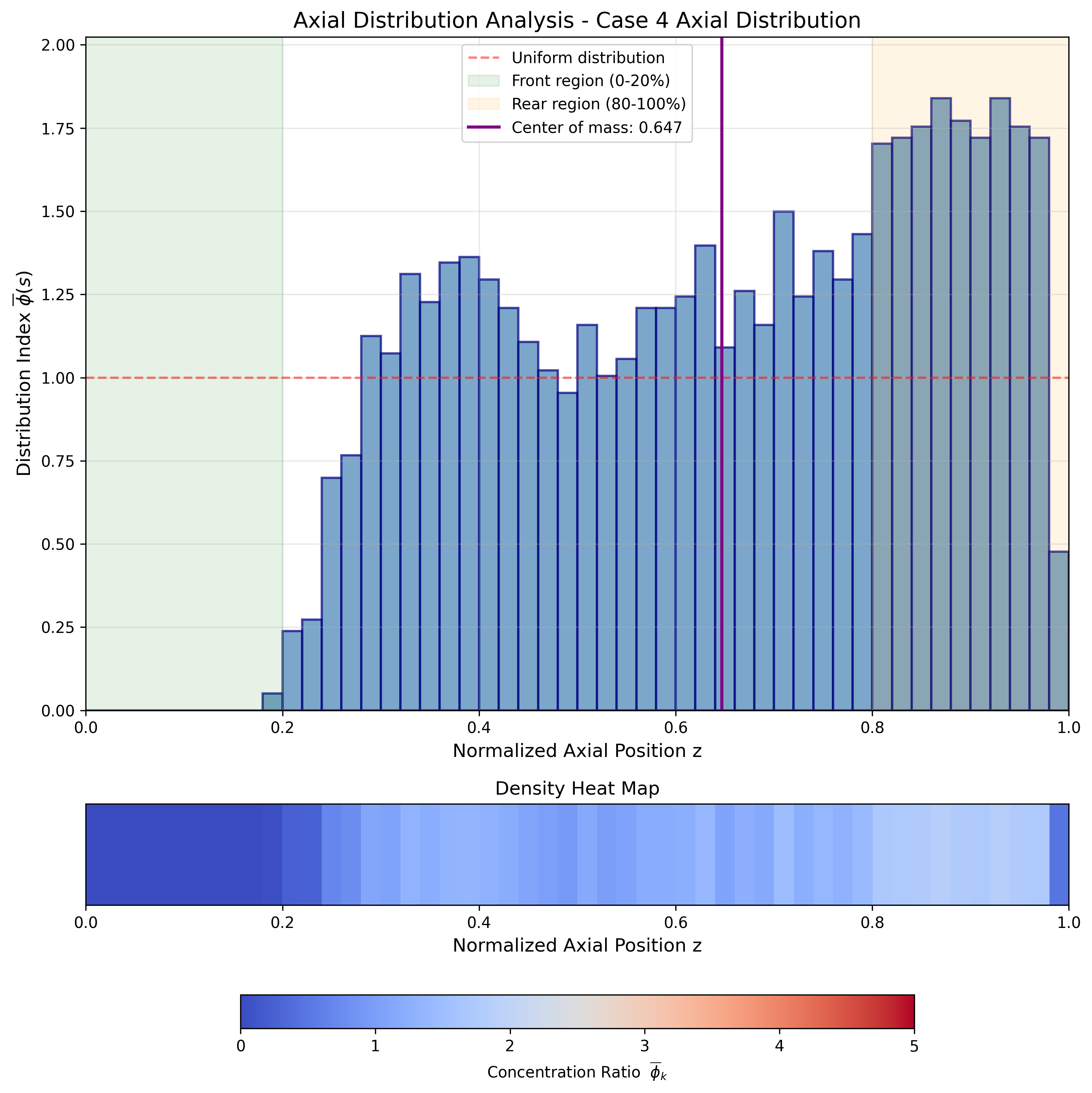}
        \caption{Case C4}
    \end{subfigure}
    \caption{Axial particle distribution $\overline{\phi}(z)$ for cases C1–C4.}
    \label{fig:axial_distribution_1_4}
\end{figure}

% Next four cases
\begin{figure}[H]
    \centering
    \begin{subfigure}[b]{0.48\textwidth}
        \includegraphics[width=\textwidth]{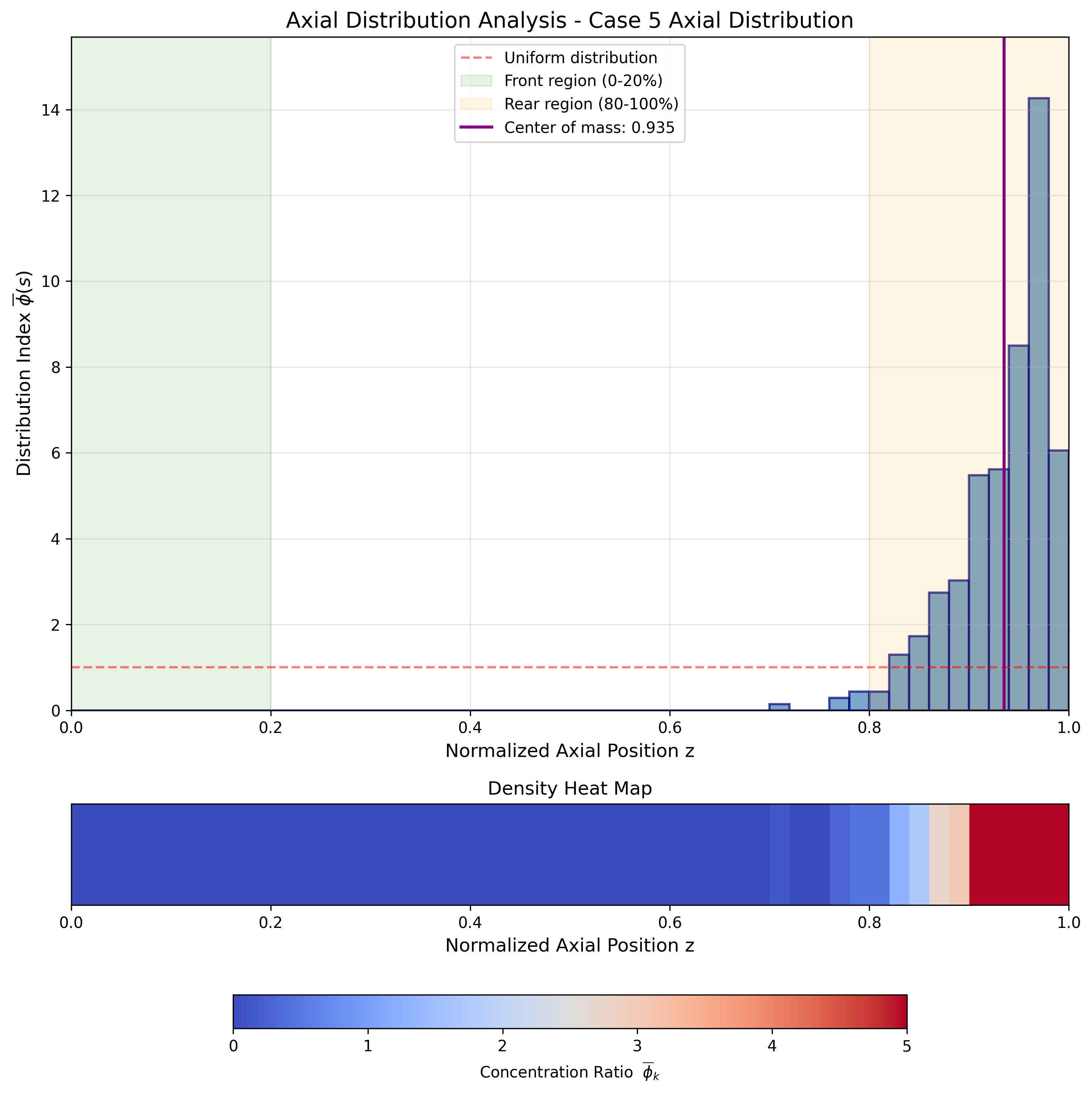}
        \caption{Case C5}
    \end{subfigure}\hfill
    \begin{subfigure}[b]{0.48\textwidth}
        \includegraphics[width=\textwidth]{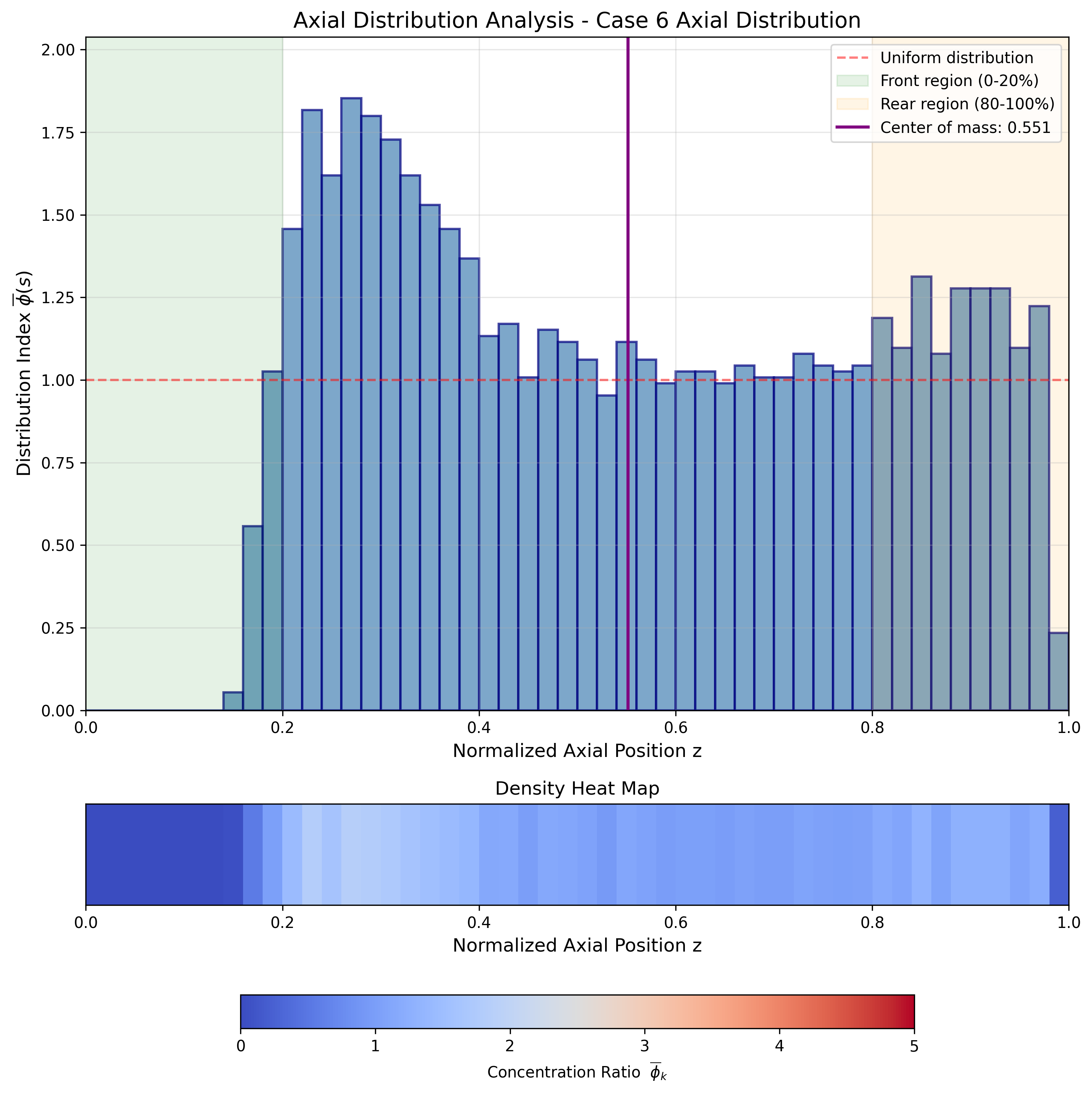}
        \caption{Case C6}
    \end{subfigure}

    \vspace{1em}

    \begin{subfigure}[b]{0.48\textwidth}
        \includegraphics[width=\textwidth]{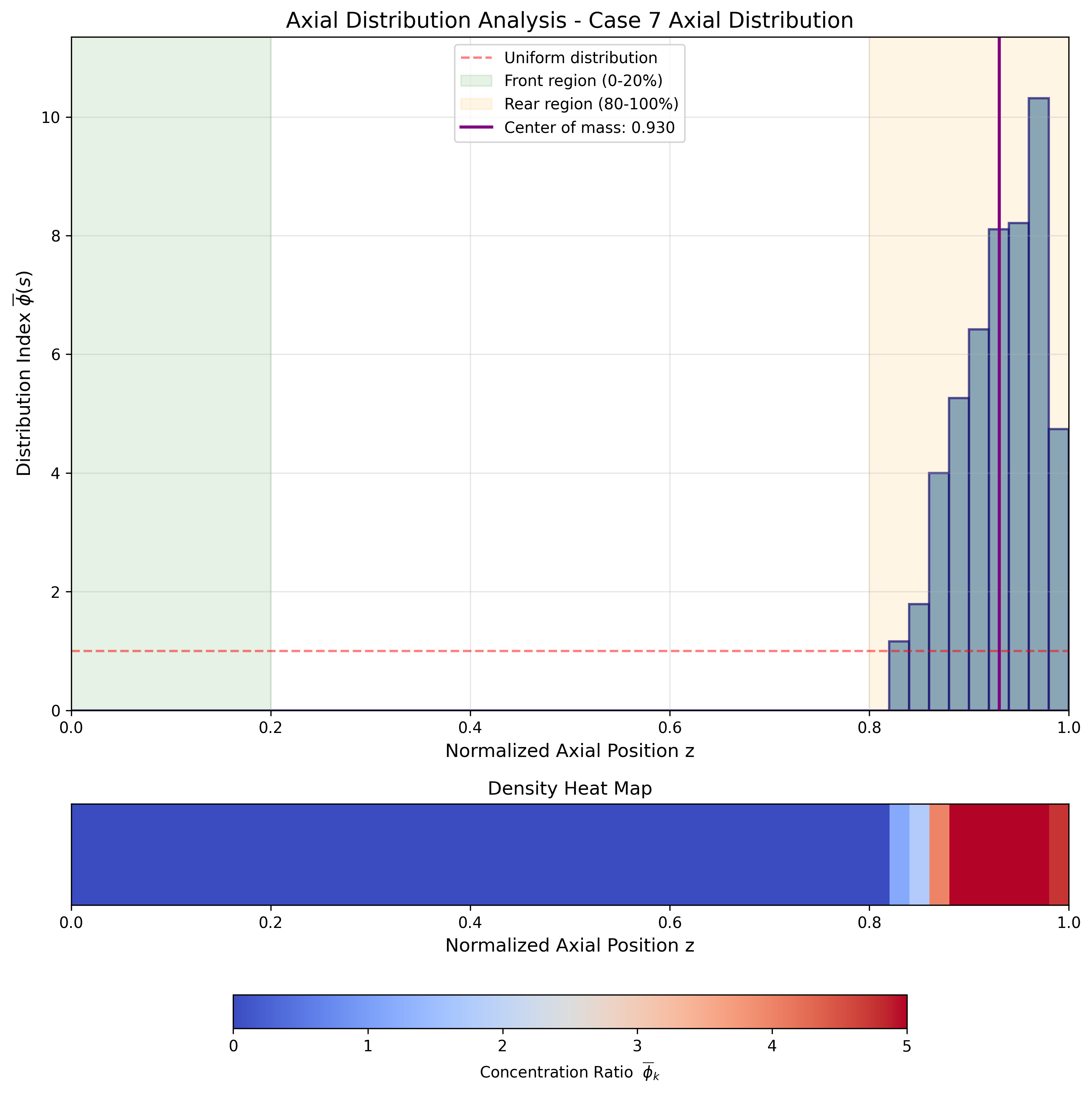}
        \caption{Case C7}
    \end{subfigure}\hfill
    \begin{subfigure}[b]{0.48\textwidth}
        \includegraphics[width=\textwidth]{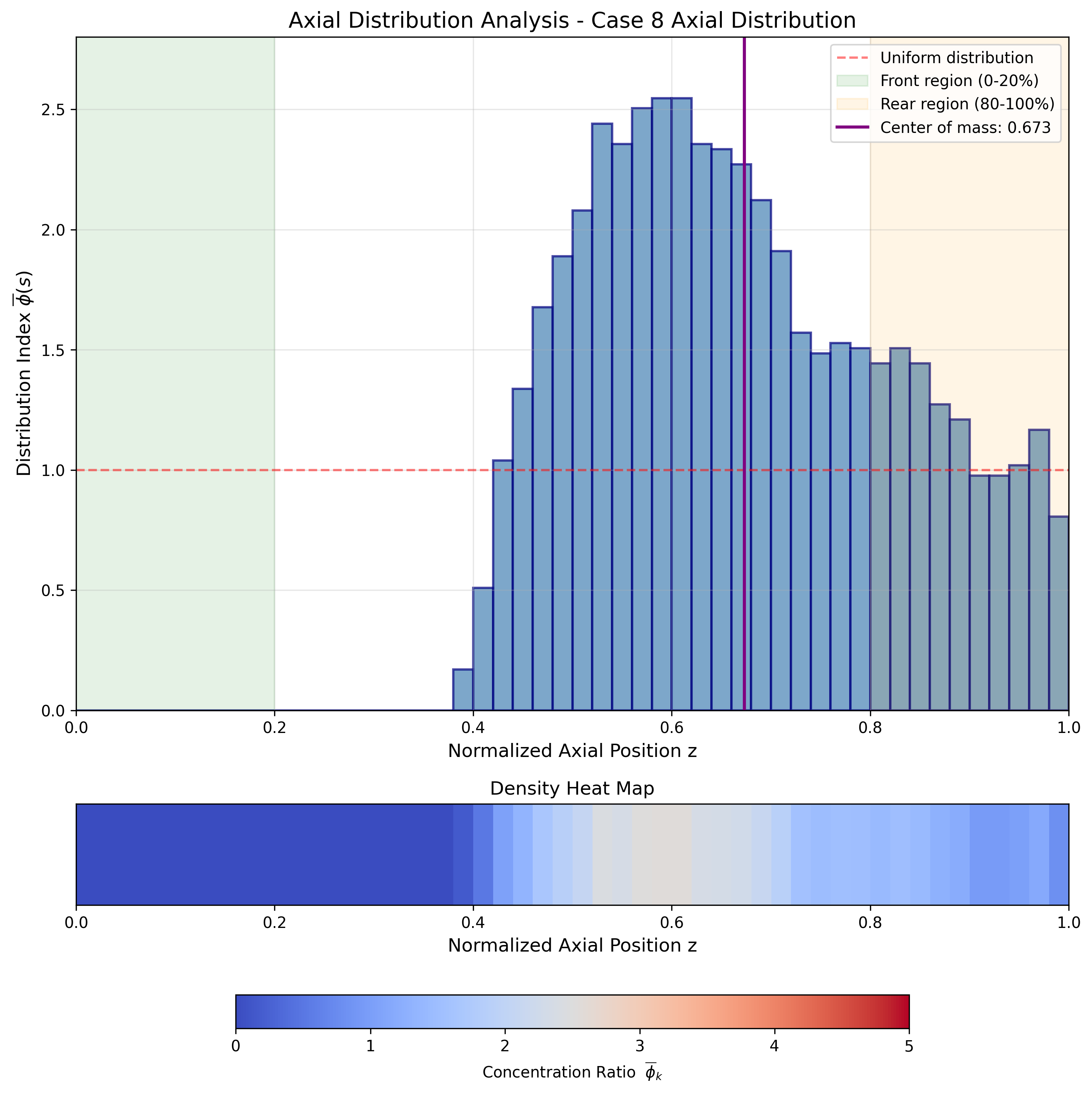}
        \caption{Case C8}
    \end{subfigure}
    \caption{Axial particle distribution $\overline{\phi}(z)$ for cases C5–C8.}
    \label{fig:axial_distribution_5_8}
\end{figure}

\clearpage
%\subsection{Radial Distribution Index $I_r$}

% First four cases
\begin{figure}[H]
    \centering
    \begin{subfigure}[b]{0.48\textwidth}
        \includegraphics[width=\textwidth]{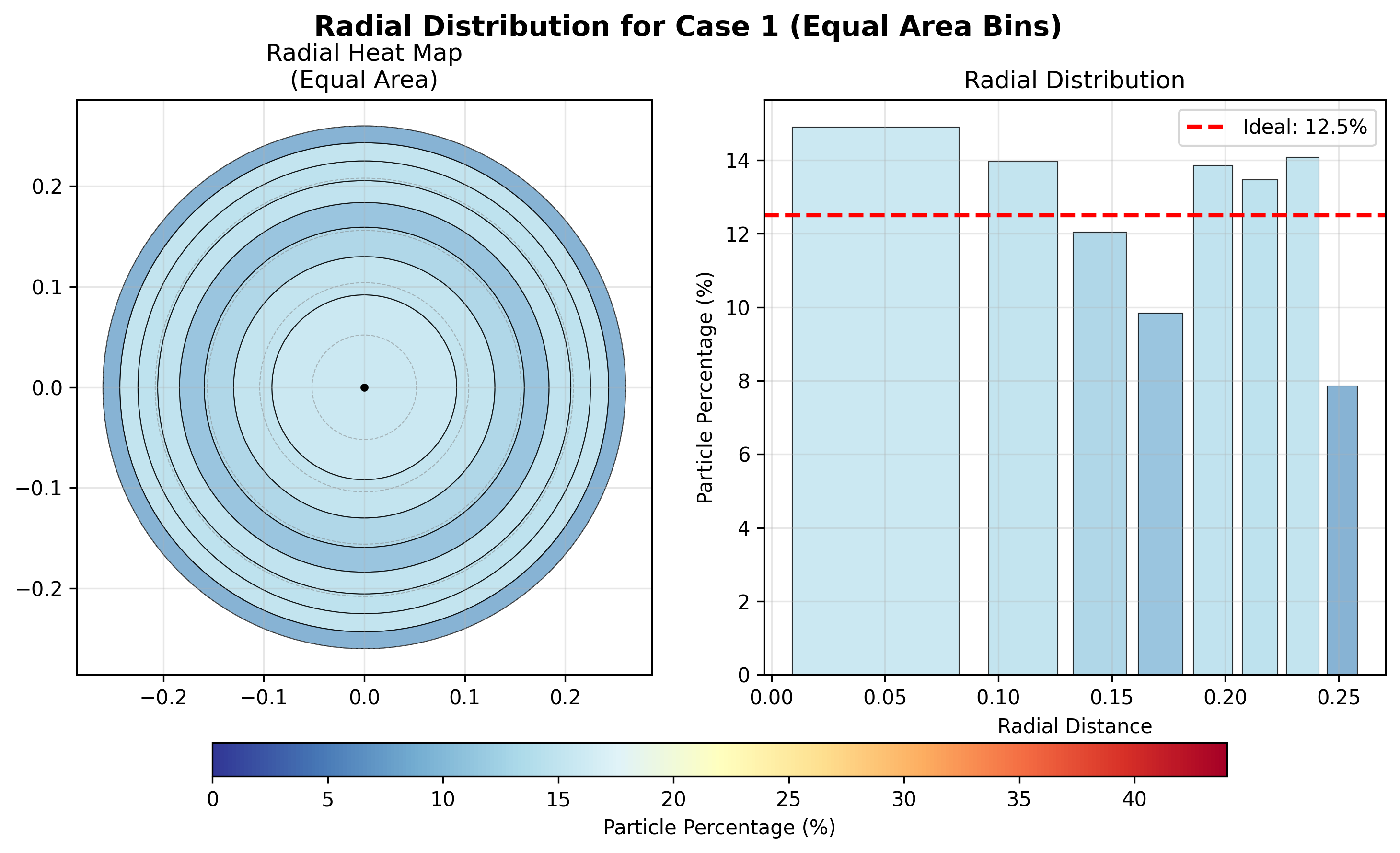}
        \caption{Case C1}
    \end{subfigure}\hfill
    \begin{subfigure}[b]{0.48\textwidth}
        \includegraphics[width=\textwidth]{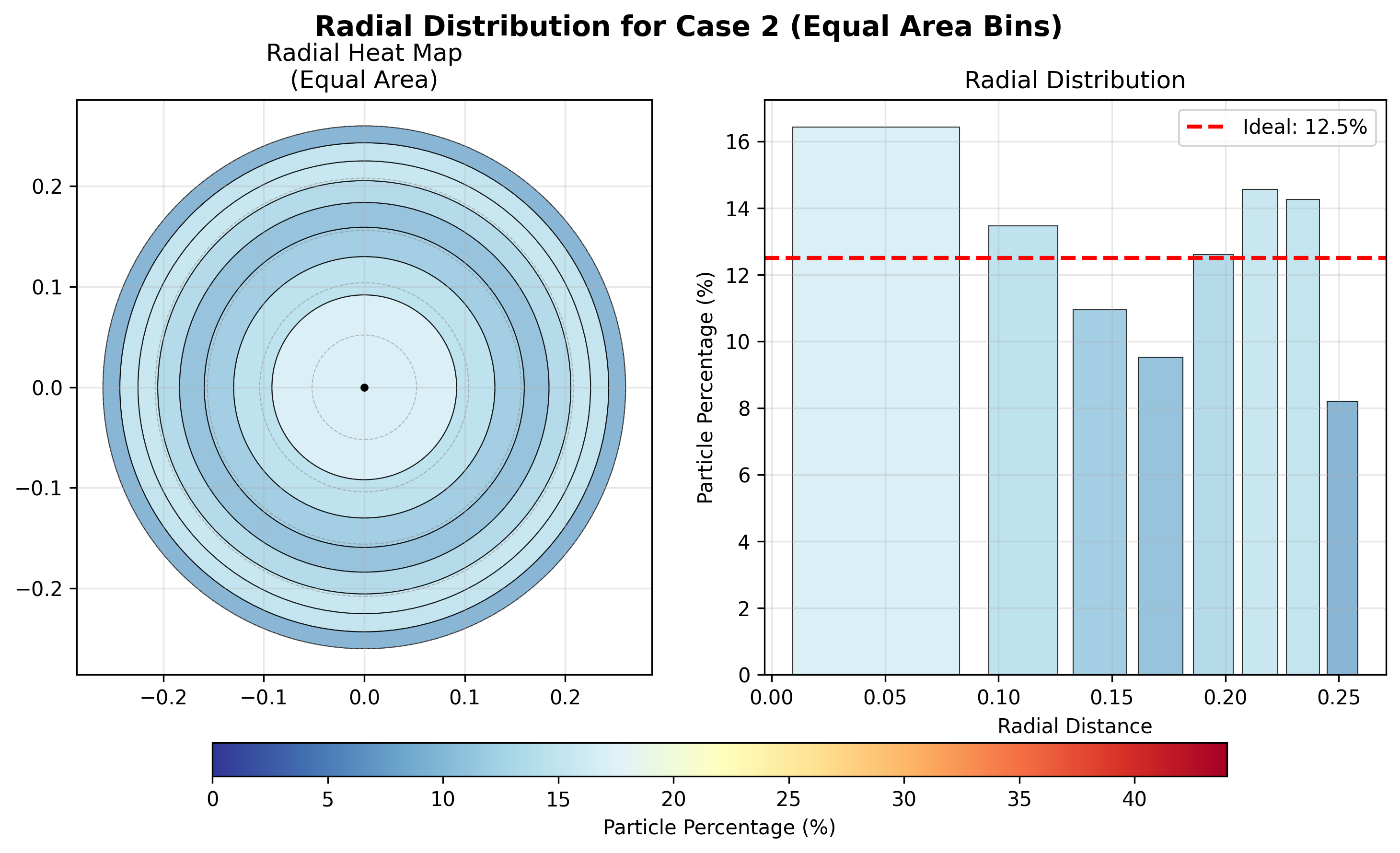}
        \caption{Case C2}
    \end{subfigure}

    \vspace{1em}

    \begin{subfigure}[b]{0.48\textwidth}
        \includegraphics[width=\textwidth]{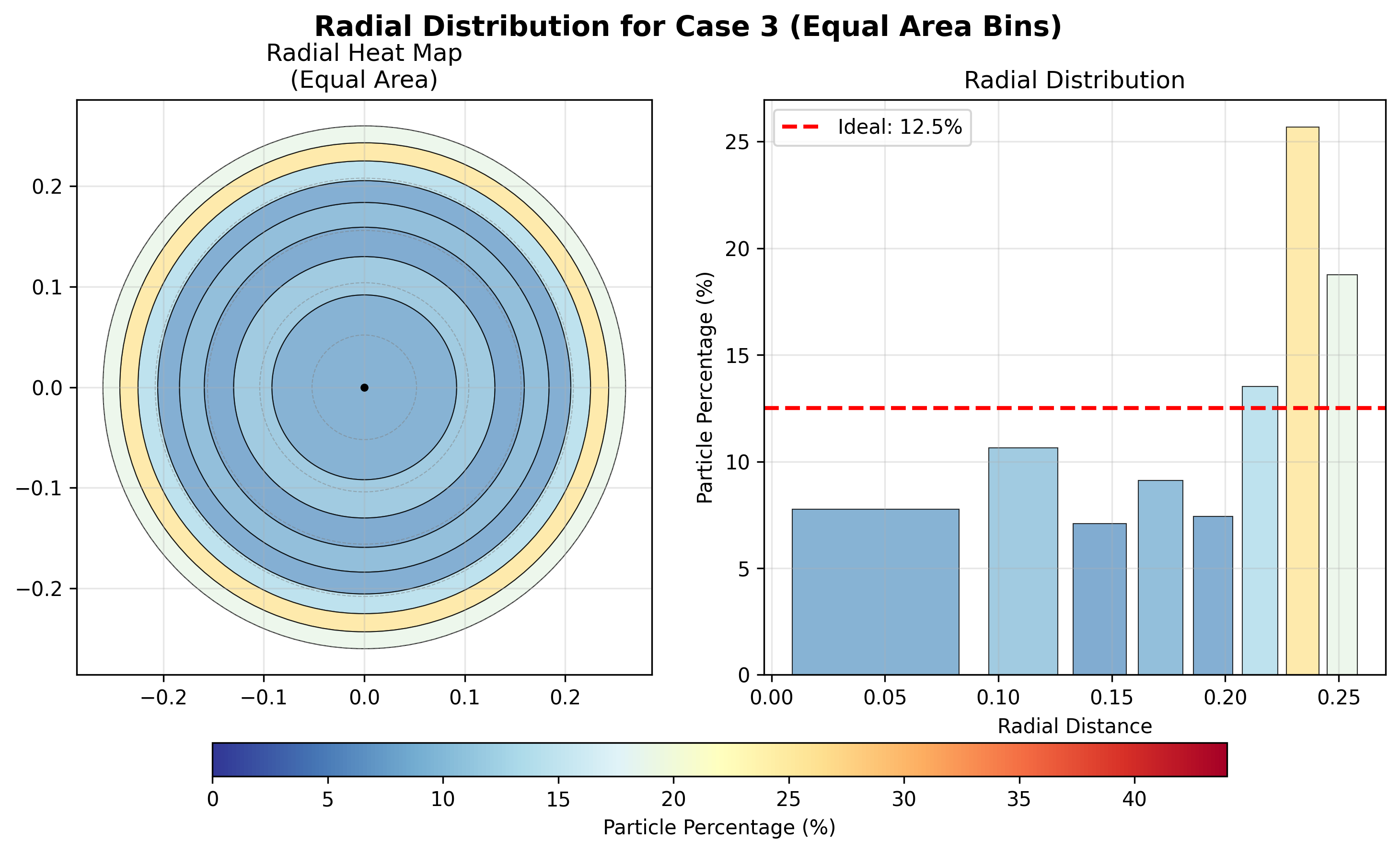}
        \caption{Case C3}
    \end{subfigure}\hfill
    \begin{subfigure}[b]{0.48\textwidth}
        \includegraphics[width=\textwidth]{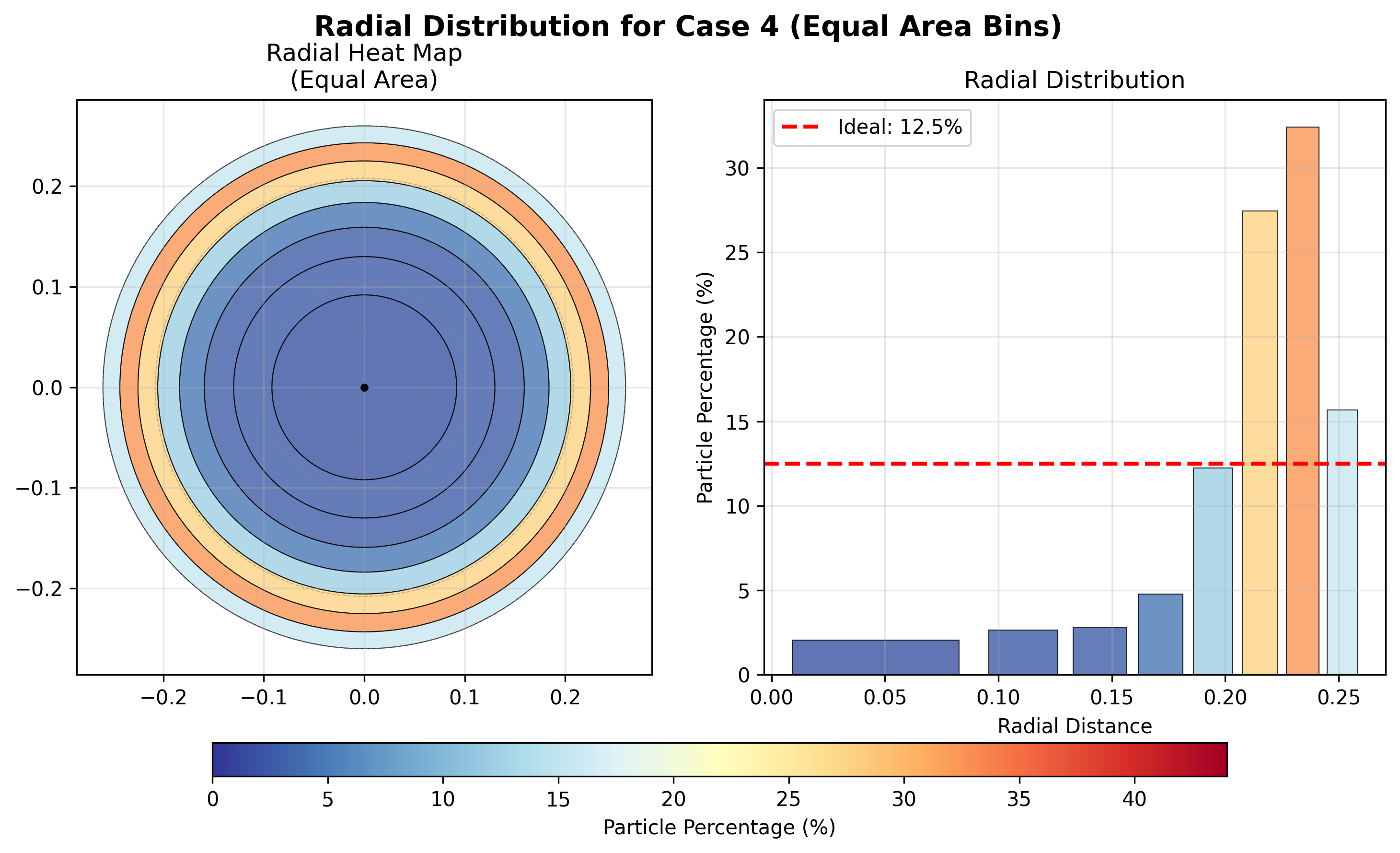}
        \caption{Case C4}
    \end{subfigure}
    \caption{Equal-area radial distribution analysis for cases C1–C4.}
    \label{fig:radial_distribution_1_4}
\end{figure}

\clearpage
% Next four cases
\begin{figure}[H]
    \centering
    \begin{subfigure}[b]{0.48\textwidth}
        \includegraphics[width=\textwidth]{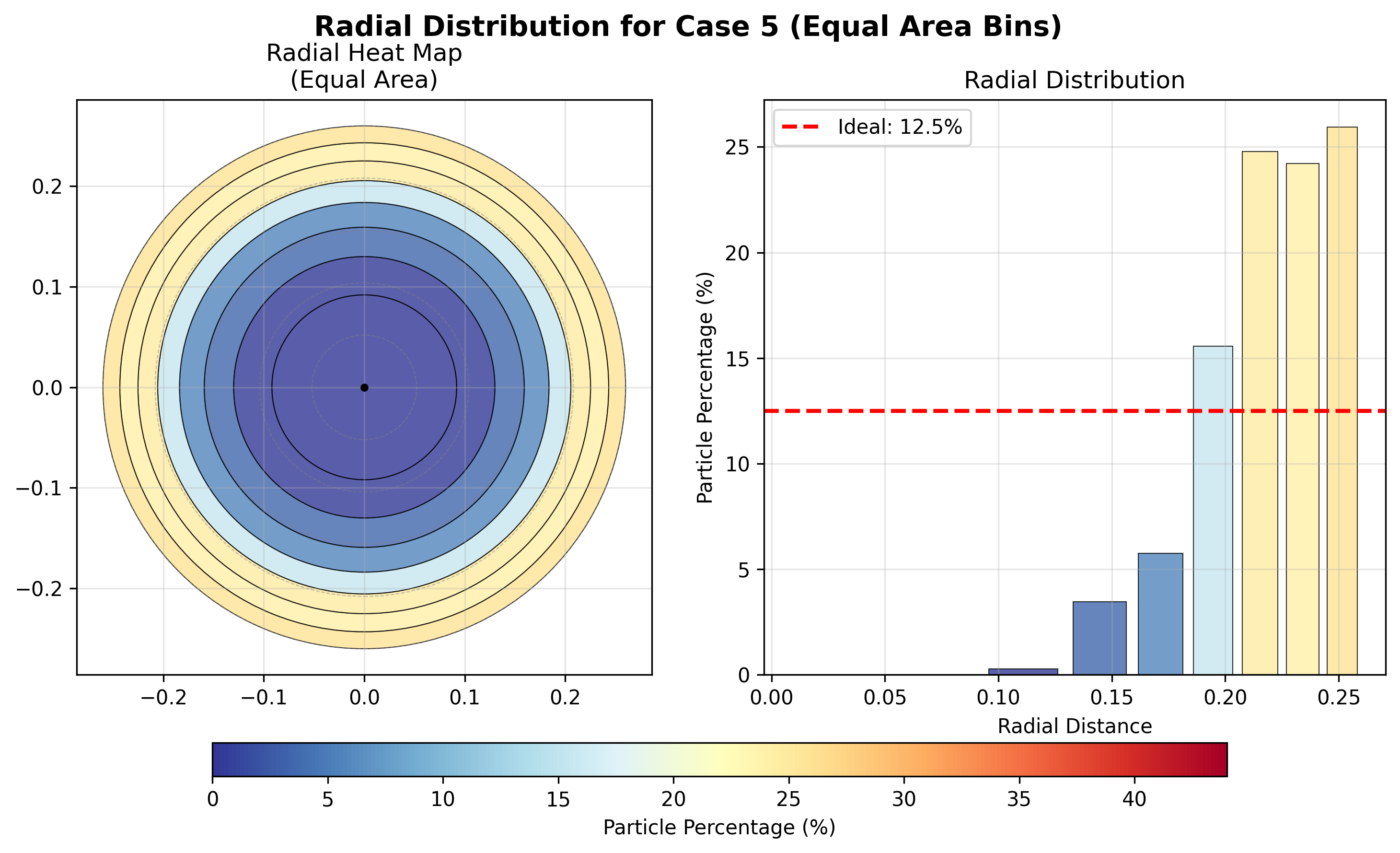}
        \caption{Case C5}
    \end{subfigure}\hfill
    \begin{subfigure}[b]{0.48\textwidth}
        \includegraphics[width=\textwidth]{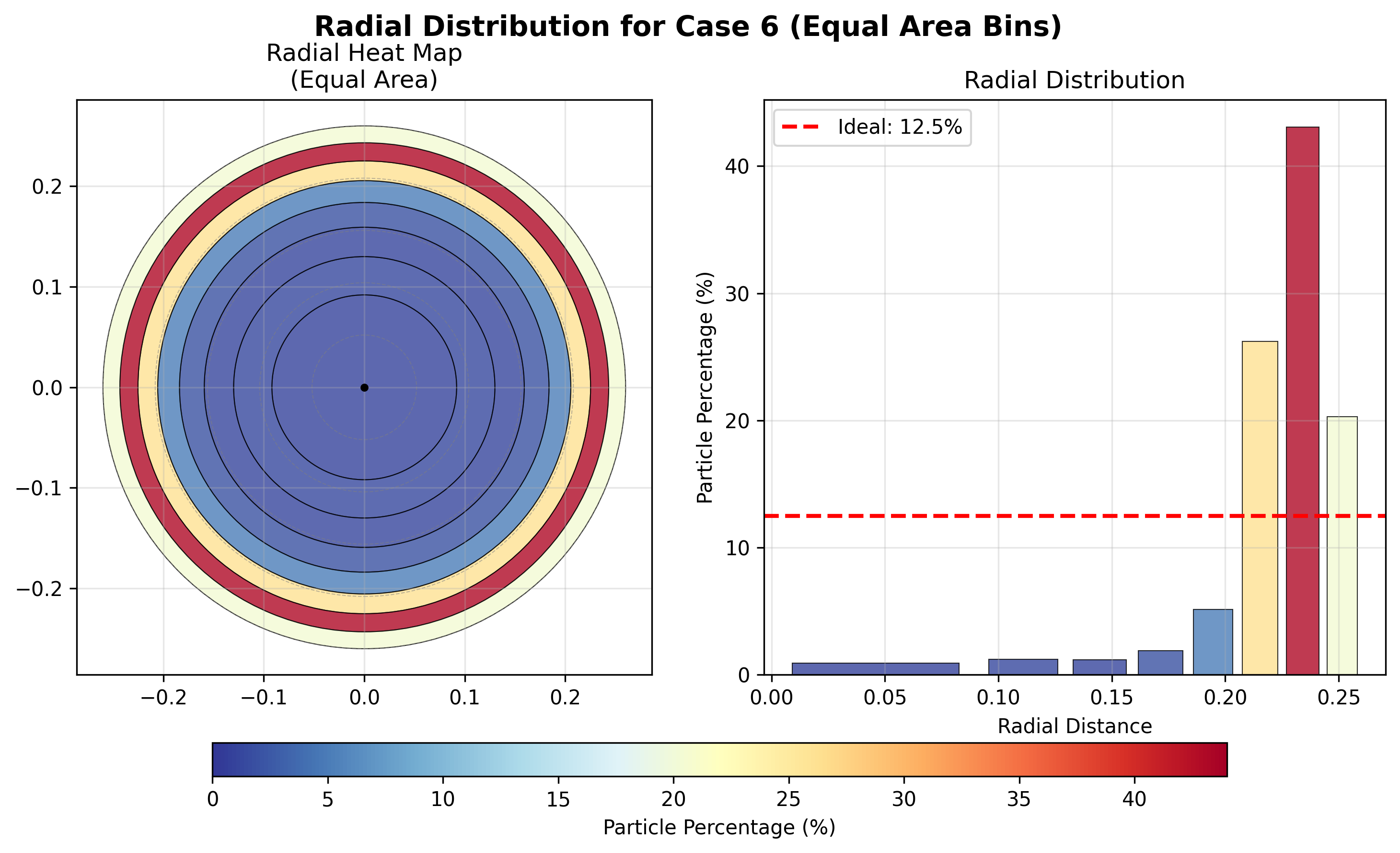}
        \caption{Case C6}
    \end{subfigure}

    \vspace{1em}

    \begin{subfigure}[b]{0.48\textwidth}
        \includegraphics[width=\textwidth]{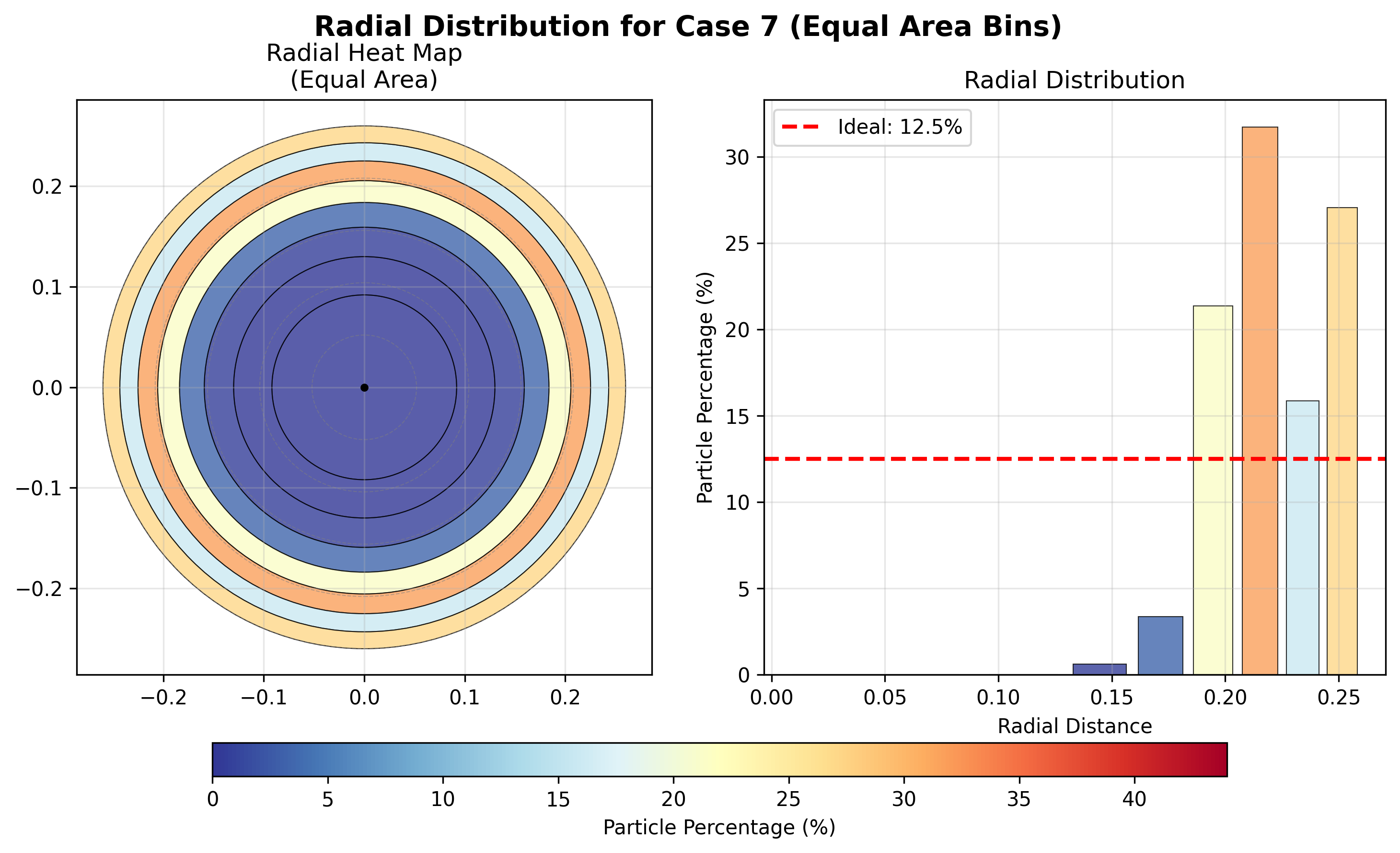}
        \caption{Case C7}
    \end{subfigure}\hfill
    \begin{subfigure}[b]{0.48\textwidth}
        \includegraphics[width=\textwidth]{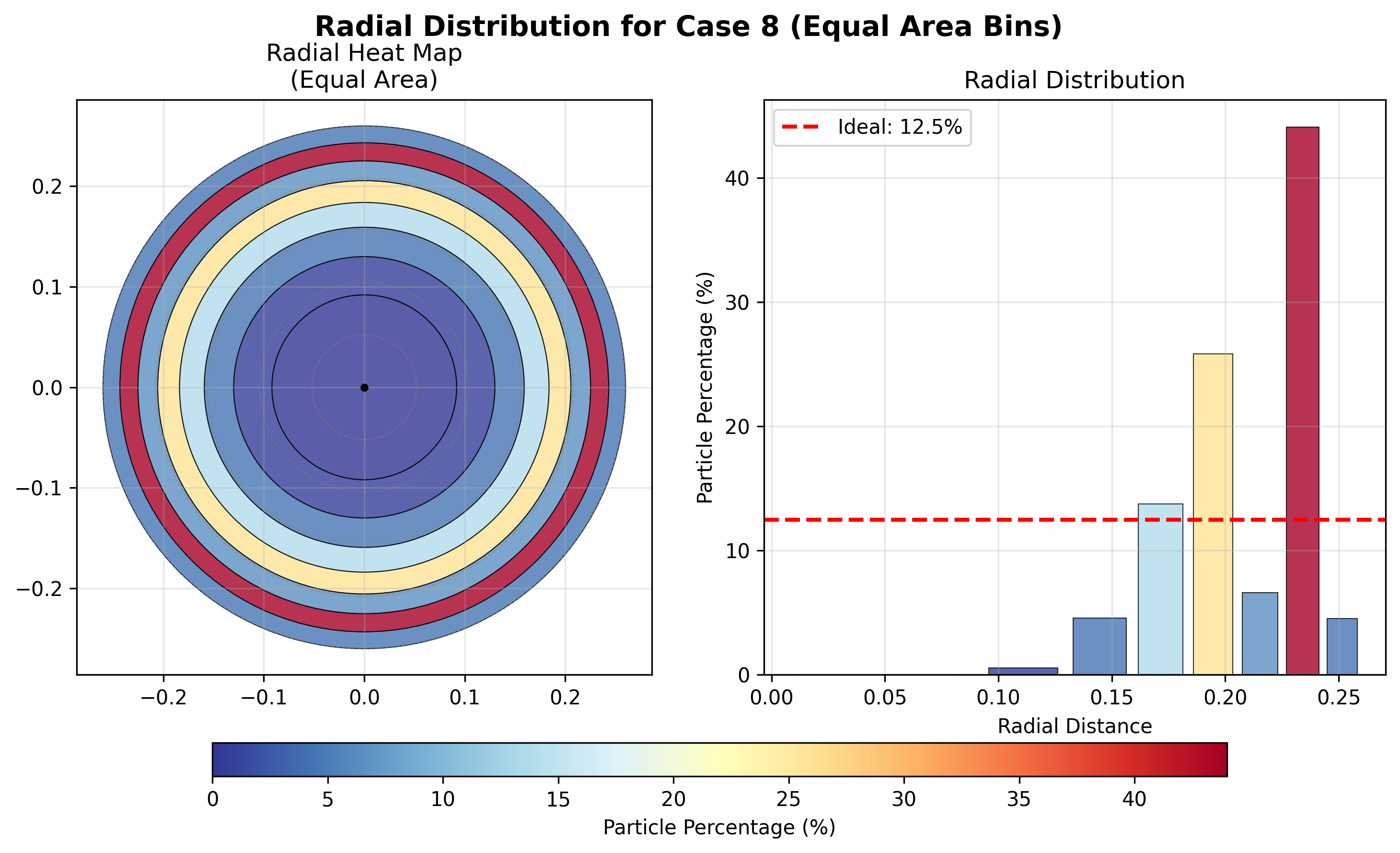}
        \caption{Case C8}
    \end{subfigure}
    \caption{Equal-area radial distribution analysis for cases C5–C8.}
    \label{fig:radial_distribution_5_8}
\end{figure}

\subsubsection{Analysis and Interpretation of Metric Evaluations}

The three quantitative descriptors---axial distribution $\phi(z)$ (see Figures \ref{fig:axial_distribution_1_4}, \ref{fig:axial_distribution_5_8}), radial index $I_r$ (see Figures \ref{fig:radial_distribution_1_4}, \ref{fig:radial_distribution_5_8}), and vertical asymmetry $A_y$---together provide a consistent picture of the suspension regimes across all simulated cases (see Table~\ref{tab:Ir_Ay_values}). By considering these metrics in combination rather than isolation, the distinct transitions between green, yellow, red/yellow, and red flow map zones can be interpreted with greater clarity.

In the \textbf{green regime} (C1--C2), the axial profiles are nearly flat, indicating homogeneous axial distributions. The radial index reaches very high values ($I_r \approx 0.90$), reflecting a uniform cross-sectional spread, while the asymmetry remains close to zero ($A_y = 0.020$ and $-0.119$). Together, these values confirm that particles are well suspended by balanced Dean vortices that counteract gravitational settling. Slight deviations in $A_y$ at higher solid content (C2) suggest mild bottom-heaviness, but radial and axial homogeneity remain largely intact.

\begin{figure} [h]
    \centering
    \includegraphics[width=0.5\textwidth]{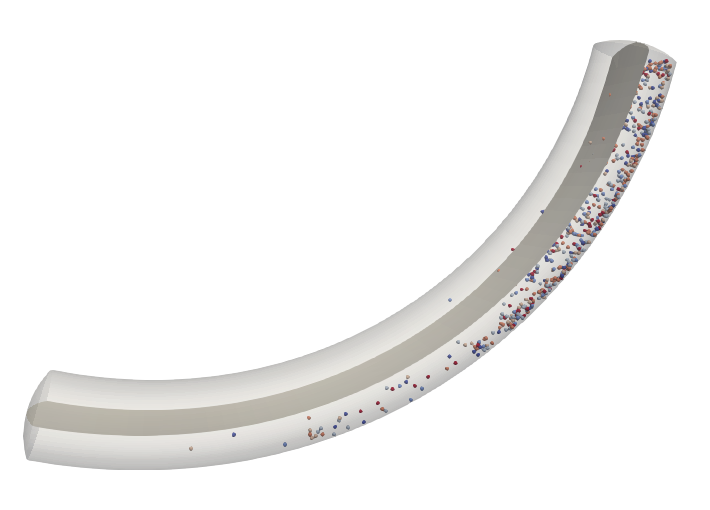}
    \caption{Fluid Domain with middle surface visualization: Even though the yellow flow regime clearly exhibits vortex-driven dynamics the influence of the vortices is not enough to populate the upper half of the domain significantly.}
    \label{fig:case3_ay}
\end{figure}

Moving into the \textbf{yellow regime} (C3--C4), the metrics shift markedly. The axial profiles show rear-loading and depleted fronts, while $I_r$ decreases to 0.77 (C3) and 0.57 (C4), indicating loss of cross-sectional uniformity. Simultaneously, $A_y$ plunges close to $-1$, confirming that particles overwhelmingly reside in the lower half of the tube (see Figure \ref{fig:case3_ay}). This joint trend reveals the collapse of upper-vortex activity and the emergence of strong gravitational bias, with only partial radial spreading persisting within the bottom half.

The \textbf{red/yellow regime} (C5--C6) continues this progression. The axial distribution becomes increasingly rear-biased, $I_r$ falls further (0.54 and 0.41), and $A_y \approx -0.95$ shows that vertical symmetry is almost entirely lost. In this state, particles not only settle toward the bottom but also accumulate strongly at the rear, consistent with stacking and poor recirculation. The middle radial bins are nearly empty, underscoring the breakdown of cross-sectional mixing once the green regime is left.

Finally, in the \textbf{red regime} (C7--C8), all three metrics confirm a gravity-dominated state. The axial profiles exhibit extreme rear loading, and $A_y = -1.000$ indicates that all particles are below the mid-plane. While the value of $I_r$ ($\approx 0.47$) appears moderate, it is misleading. It arises from geometric binning effects due to rear stacking, rather than from genuine radial uniformity. The front of the slug remains unpopulated, and the suspension effectively collapses into a dense, bottom-heavy cluster.

Taken together, these results demonstrate that meaningful upper-half particle population and front-filling occur only in the green regime. Exiting this regime leads to rapid depletion of the middle radial bins, progressive rear-loading, and strong vertical asymmetry.
Interpreting jointly $\phi(z)$, $I_r$, and $A_y$ is essential. For example, the moderately high $I_r$ in C7–C8 suggests partial uniformity, yet the combination of metrics (see Table~\ref{tab:Ir_Ay_values}) reveals the true, gravity-dominated state.  

\begin{table}[H]
\centering
\caption{Axial particle distribution $\overline{\phi}(z)$, radial distribution index $I_r$, and vertical asymmetry index $A_y$ 
for all simulated cases. A nearly flat $\overline{\phi}(z)$ profile indicates homogeneous axial distribution, while rear-loading 
or mid/rear peaks indicate segregation. Higher $I_r$ corresponds to a more uniform radial spread. 
$A_y \approx 0$ indicates vertical symmetry; $A_y < 0$ reflects bottom-heavy configurations; 
$A_y = -1$ corresponds to all particles below the mid-plane.}
\label{tab:Ir_Ay_values}
\begin{tabular}{lcccccc}
\toprule
Case & $\overline{\phi}(z)$ & $I_r$ & $A_y$ & $\phi_{\mathrm{upper}}$ & $\phi_{\mathrm{lower}}$ & Flow Map Zone \\
\midrule
C1 & nearly flat        & 0.9114 &  0.0198 & 0.5099 & 0.4901 & green \\
C2 & nearly flat        & 0.8992 & -0.1188 & 0.4406 & 0.5594 & green \\
C3 & mid–rear           & 0.7664 & -0.9764 & 0.0118 & 0.9882 & yellow \\
C4 & mid–rear           & 0.5657 & -0.8780 & 0.0610 & 0.9390 & yellow \\
C5 & rear-biased        & 0.5372 & -0.9597 & 0.0202 & 0.9798 & red/yellow \\
C6 & rear-biased        & 0.4056 & -0.9453 & 0.0273 & 0.9727 & red/yellow \\
C7 & rear-loaded        & 0.4767 & -1.0000 & 0.0000 & 1.0000 & red \\
C8 & rear-loaded        & 0.4720 & -1.0000 & 0.0000 & 1.0000 & red \\
\bottomrule
\end{tabular}
\end{table}

\section{Conclusion}

This work demonstrated the feasibility of fully resolved DNS for predicting particle suspension states in the Archimedes Tube Crystallizer (ATC). Across eight operating conditions, the simulations reproduced the experimentally established flow map with high fidelity, capturing transitions between green, yellow, red/yellow, and red regimes.  

A central outcome of this study is the introduction of three quantitative descriptors: the axial distribution $\overline{\phi}(s)$, the radial distribution index $I_r$, and the vertical asymmetry $A_y$ (Table~\ref{tab:Ir_Ay_values}). These metrics proved indispensable for identifying distinctive features of each regime, such as rear-loading, radial depletion, or symmetry loss, and for distinguishing subtle transitions that qualitative inspection alone cannot resolve. Importantly, the combined interpretation of all three avoids misleading conclusions that may arise from any single measure in isolation. Beyond validation, the metrics provide a practical diagnostic tool: the corkscrew-like geometry of the ATC makes experimental imaging difficult, while regime classification based on numerical values can be done quickly and unambiguously. The additional insights into particle distribution also have direct relevance for crystallization, where mixing, growth, and agglomeration are strongly affected by suspension quality.  

At the same time, this study strengthens the case for DNS as a method. Unlike one-way coupled or unresolved approaches, the DNS resolves momentum exchange without empirical closures, naturally captures stacking and wall-ward clustering, and remains valid in dense regimes where drag-based models fail. It thereby offers mechanistic transparency and predictive accuracy, enabling not only reliable regime classification but also the development of closures and design strategies grounded in first principles.  

Taken together, the DNS framework and the proposed metrics establish a sound and versatile foundation for ATC suspension analysis. They bridge qualitative flow maps with quantitative measures, reduce experimental ambiguity, and provide mechanistic insight that is essential for advancing crystallizer design. Future work will extend DNS to larger particle numbers, integrate it with population balance models, and further explore how these metrics can inform optimization and scale-up of the ATC concept.

\appendix
\section{Variables and Symbols}
\label{appendix:contact_symbols}

This appendix provides definitions of the symbols used in the hard-sphere contact formulation in Section~\ref{sec:lubrication}.

\begin{itemize}
    \item $\dot{\bm{g}}_i$: Relative velocity at contact $i$.
    \item $\bm{p}_i$: Contact impulse vector at contact $i$, with components along the normal and tangential directions.
    \item $\bm{W}_i$: Delassus operator (Schur complement) relating impulse to velocity change at contact $i$.
    \item $\bm{b}_i$: Bias term, including contributions from Baumgarte stabilization for constraint drift correction.
    \item $\mathcal{K}_i$: Local friction cone defining the admissible set of impulses satisfying Coulomb friction: 
    \[
    \mathcal{K}_i = \{ \bm{p}_i \mid p_n \geq 0,\; \|\bm{p}_t\| \leq \mu p_n \}.
    \]
    \item $\Pi_{\mathcal{K}_i}$: Projection operator onto the admissible set $\mathcal{K}_i$.
    \item $\omega$: Relaxation parameter in the PGS iteration, typically in $(0,1]$.
\end{itemize}

% "plain" orders items in the references list in alphabetical order
\bibliographystyle{unsrt} % or another style like abbrv, unsrt, etc. 
\bibliography{11785_project,KW,Paper_I_BibTex} % The name of your .bib file without extension

\end{document}